\documentclass[twocolumn,traditabstract]{aa}
\usepackage{graphicx,amsmath}
\usepackage{epsf}
\usepackage{txfonts}
\usepackage[hyphens]{url}
\usepackage{longtable}
\usepackage{lscape}
\usepackage[hyperindex,breaklinks=true, colorlinks, citecolor=blue]{hyperref}  
\usepackage{comment}
\usepackage{natbib}
\usepackage{upgreek}
\usepackage{float}
\usepackage[switch,pagewise]{lineno}
\usepackage{multirow}
\usepackage{caption}
\usepackage{lipsum}

\newcommand{\juan}[1]{{\color{black} #1}}

\newcommand{\commentproof}[1]{{\bf \color{green}#1}}
\renewcommand{\commentproof}[1]{} 

\providecommand{\sorthelp}[1]{}

\begin{document}

\def\setsymbol#1#2{\expandafter\def\csname #1\endcsname{#2}}
\def\getsymbol#1{\csname #1\endcsname}

\def\Planck{\textit{Planck}}

\def\HeJT{$^4$He-JT}

\def\allearlypapers{\nocite{planck2011-1.1, planck2011-1.3, planck2011-1.4, planck2011-1.5, planck2011-1.6, planck2011-1.7, planck2011-1.10, planck2011-1.10sup, planck2011-5.1a, planck2011-5.1b, planck2011-5.2a, planck2011-5.2b, planck2011-5.2c, planck2011-6.1, planck2011-6.2, planck2011-6.3a, planck2011-6.4a, planck2011-6.4b, planck2011-6.6, planck2011-7.0, planck2011-7.2, planck2011-7.3, planck2011-7.7a, planck2011-7.7b, planck2011-7.12, planck2011-7.13}}

\def\all2013resultspapers{\nocite{planck2013-p01, planck2013-p02, planck2013-p02a, planck2013-p02d, planck2013-p02b, planck2013-p03, planck2013-p03c, planck2013-p03f, planck2013-p03d, planck2013-p03e, planck2013-p01a, planck2013-p06, planck2013-p03a, planck2013-pip88, planck2013-p08, planck2013-p11, planck2013-p12, planck2013-p13, planck2013-p14, planck2013-p15, planck2013-p05b, planck2013-p17, planck2013-p09, planck2013-p09a, planck2013-p20, planck2013-p19, planck2013-pipaberration, planck2013-p05, planck2013-p05a, planck2013-pip56, planck2013-p06b, planck2013-p01a}}

\newbox\tablebox    \newdimen\tablewidth
\def\leaderfil{\leaders\hbox to 5pt{\hss.\hss}\hfil}
%
%
\def\endPlancktable{\tablewidth=\columnwidth 
    $$\hss\copy\tablebox\hss$$
    \vskip-\lastskip\vskip -2pt}
\def\endPlancktablewide{\tablewidth=\textwidth 
    $$\hss\copy\tablebox\hss$$
    \vskip-\lastskip\vskip -2pt}
\def\tablenote#1 #2\par{\begingroup \parindent=0.8em
    \abovedisplayshortskip=0pt\belowdisplayshortskip=0pt
    \noindent
    $$\hss\vbox{\hsize\tablewidth \hangindent=\parindent \hangafter=1 \noindent
    \hbox to \parindent{$^#1$\hss}\strut#2\strut\par}\hss$$
    \endgroup}
\def\doubleline{\vskip 3pt\hrule \vskip 1.5pt \hrule \vskip 5pt}

%
\def\L2{\ifmmode L_2\else $L_2$\fi}
\def\dtt{\Delta T/T}
\def\DeltaT{\ifmmode \Delta T\else $\Delta T$\fi}
\def\deltat{\ifmmode \Delta t\else $\Delta t$\fi}
\def\fknee{\ifmmode f_{\rm knee}\else $f_{\rm knee}$\fi}
\def\Fmax{\ifmmode F_{\rm max}\else $F_{\rm max}$\fi}
\def\solar{\ifmmode{\rm M}_{\mathord\odot}\else${\rm M}_{\mathord\odot}$\fi}
\def\Msolar{\ifmmode{\rm M}_{\mathord\odot}\else${\rm M}_{\mathord\odot}$\fi}
\def\Lsolar{\ifmmode{\rm L}_{\mathord\odot}\else${\rm L}_{\mathord\odot}$\fi}
\def\inv{\ifmmode^{-1}\else$^{-1}$\fi}
\def\mo{\ifmmode^{-1}\else$^{-1}$\fi}
\def\sup#1{\ifmmode ^{\rm #1}\else $^{\rm #1}$\fi}
\def\expo#1{\ifmmode \times 10^{#1}\else $\times 10^{#1}$\fi}
\def\,{\thinspace}
\def\lsim{\mathrel{\raise .4ex\hbox{\rlap{$<$}\lower 1.2ex\hbox{$\sim$}}}}
\def\gsim{\mathrel{\raise .4ex\hbox{\rlap{$>$}\lower 1.2ex\hbox{$\sim$}}}}
\let\lea=\lsim
\let\gea=\gsim
\def\simprop{\mathrel{\raise .4ex\hbox{\rlap{$\propto$}\lower 1.2ex\hbox{$\sim$}}}}
\def\deg{\ifmmode^\circ\else$^\circ$\fi}
\def\pdeg{\ifmmode $\setbox0=\hbox{$^{\circ}$}\rlap{\hskip.11\wd0 .}$^{\circ}
          \else \setbox0=\hbox{$^{\circ}$}\rlap{\hskip.11\wd0 .}$^{\circ}$\fi}
\def\arcs{\ifmmode {^{\scriptstyle\prime\prime}}
          \else $^{\scriptstyle\prime\prime}$\fi}
\def\arcm{\ifmmode {^{\scriptstyle\prime}}
          \else $^{\scriptstyle\prime}$\fi}
\newdimen\sa  \newdimen\sb
\def\parcs{\sa=.07em \sb=.03em
     \ifmmode \hbox{\rlap{.}}^{\scriptstyle\prime\kern -\sb\prime}\hbox{\kern -\sa}
     \else \rlap{.}$^{\scriptstyle\prime\kern -\sb\prime}$\kern -\sa\fi}
\def\parcm{\sa=.08em \sb=.03em
     \ifmmode \hbox{\rlap{.}\kern\sa}^{\scriptstyle\prime}\hbox{\kern-\sb}
     \else \rlap{.}\kern\sa$^{\scriptstyle\prime}$\kern-\sb\fi}
\def\ra[#1 #2 #3.#4]{#1\sup{h}#2\sup{m}#3\sup{s}\llap.#4}
\def\dec[#1 #2 #3.#4]{#1\deg#2\arcm#3\arcs\llap.#4}
\def\deco[#1 #2 #3]{#1\deg#2\arcm#3\arcs}
\def\rra[#1 #2]{#1\sup{h}#2\sup{m}}
\def\page{\vfill\eject}
\def\dots{\relax\ifmmode \ldots\else $\ldots$\fi}
%
%
\def\WHzsr{\ifmmode $W\,Hz\mo\,sr\mo$\else W\,Hz\mo\,sr\mo\fi}
\def\mHz{\ifmmode $\,mHz$\else \,mHz\fi}
\def\GHz{\ifmmode $\,GHz$\else \,GHz\fi}
\def\mKs{\ifmmode $\,mK\,s$^{1/2}\else \,mK\,s$^{1/2}$\fi}
\def\muKs{\ifmmode \,\mu$K\,s$^{1/2}\else \,$\mu$K\,s$^{1/2}$\fi}
\def\muKRJs{\ifmmode \,\mu$K$_{\rm RJ}$\,s$^{1/2}\else \,$\mu$K$_{\rm RJ}$\,s$^{1/2}$\fi}
\def\muKHz{\ifmmode \,\mu$K\,Hz$^{-1/2}\else \,$\mu$K\,Hz$^{-1/2}$\fi}
\def\MJysr{\ifmmode \,$MJy\,sr\mo$\else \,MJy\,sr\mo\fi}
\def\MJysrmK{\ifmmode \,$MJy\,sr\mo$\,mK$_{\rm CMB}\mo\else \,MJy\,sr\mo\,mK$_{\rm CMB}\mo$\fi}
\def\microns{\ifmmode \,\mu$m$\else \,$\mu$m\fi}
\def\micron{\microns}
\def\muK{\ifmmode \,\mu$K$\else \,$\mu$\hbox{K}\fi}
\def\microK{\ifmmode \,\mu$K$\else \,$\mu$\hbox{K}\fi}
\def\muW{\ifmmode \,\mu$W$\else \,$\mu$\hbox{W}\fi}
\def\kms{\ifmmode $\,km\,s$^{-1}\else \,km\,s$^{-1}$\fi}
\def\kmsMpc{\ifmmode $\,\kms\,Mpc\mo$\else \,\kms\,Mpc\mo\fi}
%
%

\providecommand{\sorthelp}[1]{}


\title{A panoptic view of the Taurus molecular cloud}
\subtitle{I. The cloud dynamics revealed by gas emission and 3D dust}
\titlerunning{Taurus I: Cloud dynamics revealed by 3D dust and atomic and molecular gas emission.}
    \author{
        J.~D.~Soler$^{1}$\thanks{Corresponding author, \email{juandiegosolerp@gmail.com}},
        C.~Zucker$^{2}$,
        J.~E.~G.~Peek$^{2}$, 
        M.~Heyer$^{3}$,
        P.~F.~Goldsmith$^{4}$,
        S.~C.~O.~Glover$^{5}$,
        S.~Molinari$^{1}$,
        R.~S.~Klessen$^{5,6}$,
        P.~Hennebelle$^{7}$,
        L.~Testi$^{8}$,
        T.~Colman$^{7}$,
        M.~Benedettini$^{1}$,
        D.~Elia$^{1}$,
        C.~Mininni$^{1}$,
        S.~Pezzuto$^{1}$,
        E.~Schisano$^{1}$,
        A.~Traficante$^{1}$
} 
\institute{
1. Istituto di Astrofisica e Planetologia Spaziali (IAPS). INAF. Via Fosso del Cavaliere 100, 00133 Roma, Italy.\\
2. Space Telescope Science Institute, 3700 San Martin Drive, Baltimore, MD 21218, USA.\\
3. Astronomy Department, University of Massachusetts, Amherst, MA, 01003, USA.\\
4. Jet Propulsion Laboratory, California Institute of Technology, 4800 Oak Grove Drive, Pasadena, CA 91109, USA.\\
5. Universit\"{a}t Heidelberg, Zentrum f\"{u}r Astronomie, Institut f\"{u}r Theoretische Astrophysik, Albert-Ueberle-Str. 2, 69120, Heidelberg, Germany.\\
6. Universit\"{a}t Heidelberg, Interdiszipli\"{a}res Zentrum f\"{u}r Wissenschaftliches Rechnen, 69120 Heidelberg, Germany.\\
7. Laboratoire AIM, Paris-Saclay, CEA/IRFU/SAp - CNRS - Universit\'{e} Paris Diderot. 91191, Gif-sur-Yvette Cedex, France.\\
8. Dipartimento di Fisica e Astronomia, Universit\`{a} di Bologna, Via Gobetti 93/2, 40122 Bologna, Italy.
}
\authorrunning{Soler,\,J.D. et al.}

\date{Received 24FEB2023 / Accepted 22MAY2023}

\abstract{
We present a study of the three-dimensional (3D) distribution of interstellar dust derived from stellar extinction observations toward the Taurus molecular cloud (MC) and its relation with the neutral atomic hydrogen (H{\sc i}) emission at 21\,cm wavelength and the carbon monoxide $^{12}$CO and $^{13}$CO emission in the $J$\,$=$\,$1$\,$\rightarrow$\,$0$ transition.
We used the histogram of oriented gradients (HOG) method to match the morphology in a 3D reconstruction of the dust density (3D dust) and the distribution of the gas tracers' emission.
The result of the HOG analysis is a map of the relationship between the distances and radial velocities.
The HOG comparison between the 3D dust and the H{\sc i} emission indicates a morphological match at the distance of Taurus but an anti-correlation between the dust density and the H{\sc i} emission, which uncovers a significant amount of cold H{\sc i} within the Taurus MC.
The HOG between the 3D dust and $^{12}$CO reveals a pattern in radial velocities and distances that is consistent with converging motions of the gas in the Taurus MC, with the near side of the cloud moving at higher velocities and the far side moving at lower velocities. 
This convergence of flows is likely triggered by the large-scale gas compression caused by the interaction of the Local Bubble and the Per-Tau shell, with Taurus lying at the intersection of the two bubble surfaces.}

\maketitle

\section{Introduction}\label{section:introduction}

\juan{Most stars form in a cold, dense, and mostly molecular phase of the interstellar medium \citep[ISM; see,][for a review]{ferriere2001,klessenANDglover2016}.
Although there is a significant amount of diffuse molecular gas in the Milky Way and other galaxies \cite[see, for example,][]{roman-duval2016,miville-deschenes2017,rosolowsky2021}, most of the star formation appears related to localized dense molecular clouds \citep[MCs; see,][for a review]{heyerANDdame2015,dobbs2014,chevance2022}.}
\juan{Thus,} insight into the formation and destruction of MCs is crucial for understanding the star formation process and the evolution of galaxies.
In this paper, we combined the atomic and molecular gas emission with the dust 3D distribution in and around the Taurus MC in search of clues about its formation and current dynamics.

At a distance of \juan{around} 140\,pc \citep[][]{zucker2020}, Taurus is one of the nearest and best-studied star-forming regions \citep[see][for a review]{kenyon2008}.
It was first discovered as a series of dark spots and lanes visible against the bright background of stars just south of the Milky Way's plane \citep{barnard1919,lynds1962} and is one of the MCs where the relation between dust and gas was first studied \citep{lilley1955}.
This region does not contain luminous O or B stars but has a rich population of less-massive pre-main sequence stars, among them T Tauri variable stars with G, K, or M
spectral types \citep{joy1945,kenyon1995,rebull2010}.

The first large-scale and high angular resolution study resolution of Taurus in carbon monoxide (CO) emission in the $J$\,$=$\,$1$\,$\rightarrow$\,$0$ transition revealed a mass of 2.4\,$\times$\,10$^{4}$\,$M_{\odot}$ in molecular gas, which in combination with the counts of young stars results in a star formation efficiency between 0.3\%\ and 1.2\%\ \citep{goldsmith2008}.
Observations of neutral atomic hydrogen (H{\sc i}) emission toward dark clouds in Taurus reveal the presence H{\sc i} narrow self-absorption (HINSA) features that appear correlated with the hydroxyl radical (OH) and CO emission, which suggests that a significant fraction of the H{\sc i} is located in the cold, well-shielded portions of MCs and is mixed with the molecular gas \citep{li2003,li2005}.

Three-dimensional maps of the local dust distribution in 3D indicate that the Taurus MC sits in the nearest hemisphere of a near-spherical shell with a diameter of around 160\,pc that is also connected to the Perseus MC, thus called the Per-Tau Shell \citep{leike2020,bialy2021}.
The connection between the Taurus and Perseus MCs had been suggested since the first large-scale mapping of the CO emission toward these regions \citep{ungerechts1987}. 
Still, it was only confirmed with the advent of the stellar parallax measurements from ESA's {\it Gaia} satellite, which along with other observations, enable the mapping of dust in 3D \citep[see, for example,][]{lallement2018,green2019}.
The Per-Tau Shell is most likely the result of stellar and supernova (SN) feedback that swept up the surrounding interstellar medium (ISM), leading to the accumulation of matter that formed the Perseus and Taurus MCs.
Using the observations of the soft X-ray continuum and the $^{26}$Al line emission, \cite{bialy2021} estimate that the Per-Tau Shell's age is between 6 and 22\,Myrs. 

Recent studies presented in \cite{zucker2022} reveal that in addition to touching the near side of the Per-Tau shell, Taurus lies at the edge of the Local Bubble.
The Local Bubble is the cavity of the low-density, high-temperature plasma surrounded by a shell of cold, neutral gas and dust in which the Sun lies \citep[see, for example,][]{sanders1977,coxANDreynolds1987,snowden2015,pelgrims2020}.
Its expansion was most likely powered by 15 to 20 SNe that have exploded over the past 14\,Myr \citep{zucker2022}. 
Thus, Taurus's position, nestled at the intersection of two shells, suggests that it may have formed by the material gathered in the collision of these two bubbles.

In this paper, we focus on the current dynamics of the Taurus molecular cloud using new data and a new method.
We used the new 3D dust maps obtained with a hierarchical Bayesian model based on stellar parallax and extinction observations from {\it Gaia} and other observatories \citep{leike2019, leike2020}.
We employed the new histogram of oriented gradients \juan{\citep[HOG;][]{soler2019a}}
method to identify the correlation in the extended emission from the different tracers.
We applied the HOG method to connect the distribution of the 3D dust map and the H{\sc i} and CO emission toward the Taurus MC.
\juan{The result is a map of the relationship between} distances and radial velocities \juan{that we used} to elucidate the MC dynamics.
Radio emission from the ISM and 3D dust have previously been combined to investigate the ISM in four dimensions \citep[three spatial and one velocity;][]{tchernyshyov2017}, but never at the level of distance precision we use here, and never before with the HOG.

This manuscript is organized as follows.
We present the H{\sc i}, CO, and 3D dust observations in Sec.~\ref{sec:data}
In Sec.~\ref{sec:methods}, we summarize the main aspects of the {\tt astroHOG} method and introduce the circular statistical tools used to quantify the correlation between the observations.
Section~\ref{sec:results} presents the results of the comparison between the gas tracers and the 3D dust model.
We present a discussion of our results and \juan{their} implication for the understanding of the Taurus region and other MCs in Sec.~\ref{sec:discussion}.
Finally, we present our conclusions in Sec.~\ref{sec:conclusions}.
We reserve complementary analysis to a set of appendices.
Appendix~\ref{appendix:significance} presents the tests performed to evaluate the significance of the HOG results.
\juan{Appendix~\ref{appendix:smoothing} shows the results of different smoothing procedures and masking procedures in the computation of the HOG.}
In Appendix~\ref{appendix:masking}, we present the HOG results obtained when applying column density masks to study the morphological correlation within the extinction-based boundaries assigned to the Taurus MC.
We present the HOG results obtained when dividing the region into blocks in Appendix~\ref{appendix:blocks}.
Finally, Appendix~\ref{appendix:extinctionmaps} presents the results of the HOG correlation between the gas tracers and the 3D extinction maps presented in \cite{green2019}.

\section{Data}\label{sec:data}

\begin{figure*}[ht!]
\centerline{
\includegraphics[width=0.495\textwidth,angle=0,origin=c]{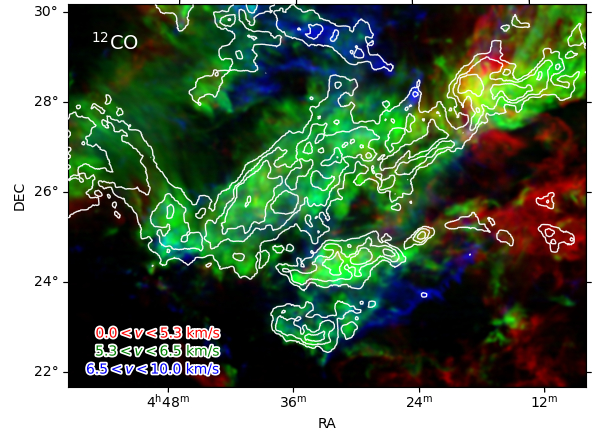}
\includegraphics[width=0.495\textwidth,angle=0,origin=c]{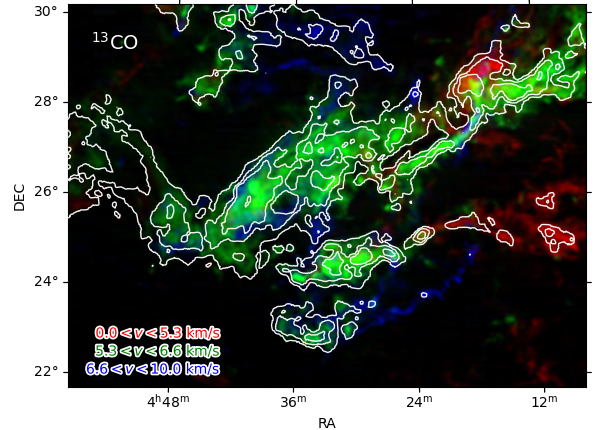}
}
\vspace{4.0mm}
\centerline{
\includegraphics[width=0.495\textwidth,angle=0,origin=c]{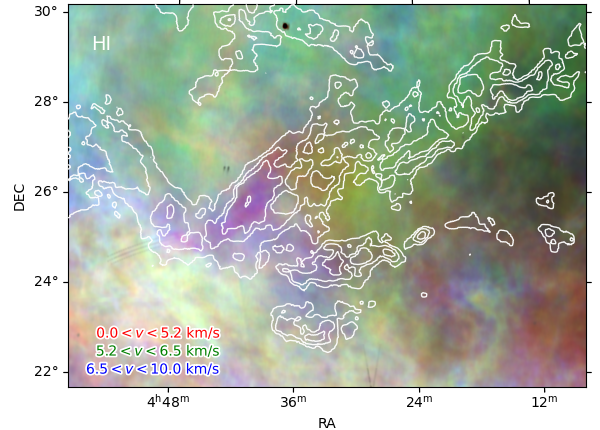}
\includegraphics[width=0.495\textwidth,angle=0,origin=c]{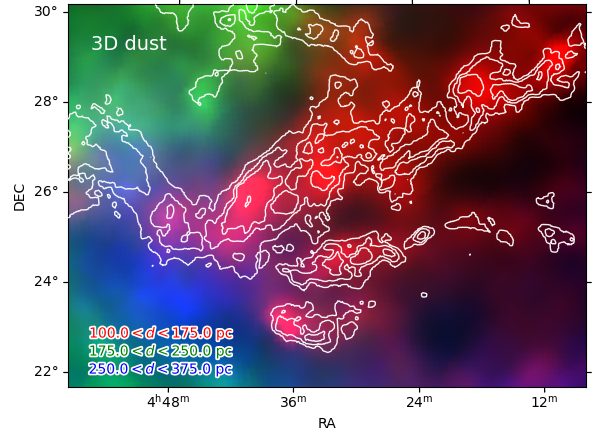}
}
\vspace{4.0mm}
\centerline{
\includegraphics[width=0.583\textwidth,angle=0,origin=c]{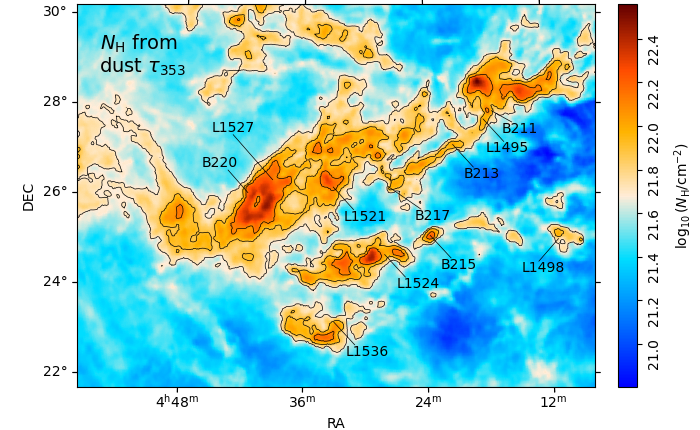}
}
\caption{The Taurus molecular cloud as seen in carbon monoxide, $^{12}$CO($J$\,$=$\,$1$\,$\rightarrow$\,$0$) and $^{13}$CO($J$\,$=$\,$1$\,$\rightarrow$\,$0$) emission from \cite{goldsmith2008}, neutral atomic hydrogen (H{\sc i}) emission from GALFA-H{\sc i} DR2 \citep{peek2018}, volume density of hydrogen nuclei ($n_{\rm H}$) across distances derived from the 3D dust reconstruction in \cite{leike2020}, \juan{and hydrogen nuclei column density ($N_{\rm H}$) derived from the dust optical depth at 353\,GHz ($\tau_{353}$) obtained with the {\it Planck} observations \citep{planck2013-p06b}.}
The contours correspond to $N_{\rm H}$\,$=$\,5\,$\times$\,$10^{21}$, 8\,$\times$\,$10^{21}$, and 1.25\,$\times$\,$10^{22}$\,cm$^{-2}$.
\juan{The labels in the bottom panel indicate the reference positions for the most prominent objects in Lynds' Catalogue of Dark Nebulae \citep{lynds1962} and Barnard's Catalogue of Dark Objects in the Sky \citep{barnard1927}.}
}
\label{fig:HIandCOand3Ddust_maps}
\end{figure*}

\subsection{Carbon monoxide (CO) emission}

We used the 100\,deg$^{2}$ survey of the Taurus molecular cloud region in $^{12}$CO and $^{13}$CO emission in the $J$\,$=$\,1\,$\rightarrow$\,0 transition presented in \cite{goldsmith2008}.
These data were obtained using the 32-pixel SEQUOIA focal plane array receiver installed in the 13.7-m Quabbin millimeter-wave telescope of the Five College Radio Astronomy Observatory \citep{erickson1999}.
The full width at half maximum (FWHM) beam sizes of the telescope are 45\arcsec\ and 47\arcsec\ for the frequencies of 115.271202 and 110.201353\,GHz, which correspond to the emission lines $^{12}$CO$(1$\,$\rightarrow$\,$0)$ and $^{13}$CO$(1$\,$\rightarrow$\,$0)$, respectively.
Further details of the data taking, data reduction, and calibration procedures are presented in \cite{narayanan2008}.

The final emission cubes of both isotopologues are very close to the Nyquist sampled and are presented in a uniform grid of 20\arcsec\ spacing, with 2069 pixels in right ascension (RA) and 1529 pixels in declination (DEC).
In the spectral axis, the data covers the radial velocity range between $-5$ and $14.9$\,km\,s$^{-1}$ in 80 spectral channels for $^{12}$CO and 76 channels for $^{13}$CO, which correspond to channel widths of 0.26 and 0.27\,km\,s$^{-1}$.
For these channel widths, the mean root-mean-square antenna temperatures in this data set are 0.28\,K and 0.125\,K for $^{12}$CO and $^{13}$CO, respectively.

\subsection{Atomic hydrogen (H{\sc i}) emission}

We used the publicly-available observations from the Galactic Arecibo L-Band Feed Array H{\sc i} survey (GALFA-H{\sc i}) data release 2 \citep[DR2,][]{peek2018}.
This survey covers the H{\sc i} emission from $-650$ to $650$\,km\,s$^{-1}$, with 0.184\,km\,s$^{-1}$ channel spacing, 4\arcmin\ angular resolution, and 150\,mK  root mean square (rms) noise per 1\,km\,s$^{-1}$ velocity channel.
The GALFA-H{\sc i} DR2 observations cover the entirety of the sky available from the William E. Gordon 305\,m antenna at Arecibo, from $\delta$\,$=$\,$-1$\deg17\arcmin\ to $+37$\deg57\arcmin\ across the whole right ascension ($\alpha$) range.
This data set provides the highest-resolution map of extended H{\sc i} emission toward Taurus available to date.

We obtained the GALFA-H{\sc i} DR2 distributed in FITS-format cubes in the ``narrow'' velocity range ($v$\,$\leq$188\,km\,s$^{-1}$). through the GALFA collaboration website\footnote{\url{https://purcell.ssl.berkeley.edu/}}.
We used the Python {\tt spectral-cube} and {\tt reproject\footnote{\url{https://reproject.readthedocs.io}}} packages to arrange these cubes into a 20\arcsec\ grid matching the coverage of the $^{12}$CO and $^{13}$CO data, but maintaining the native spectral axis with a spectral resolution of 0.184\,km\,s$^{-1}$.
This operation implies oversampling the 4\arcmin\ beam, but we accounted for this fact using the statistical weights introduced in Sec.~\ref{sec:methods}.

\subsection{Dust}

\subsubsection{3D dust}

We used the 3D dust distribution cube presented in \cite{leike2020}.
This model has a higher spatial resolution than other state-of-the-art 3D dust reconstructions \citep[for example,][]{rezaei2022,vergely2022,dharmawardena2023}.
Since the statistical significance of the HOG analysis depends on the number of independent gradients in the compared images, this advantage is crucial for the study presented in this paper.

The \cite{leike2020} dust cube was derived from stellar distances and integrated extinction estimates from the Starhorse catalog \citep{anders2019}, which combined observations from ESA's {\it Gaia} satellite second data release \citep[DR2;][]{gaia2018A&A...616A...1G}, NASA's Wide-field Infrared Survey Explorer (WISE) mission \citep{wright2010}, the Two Micron All Sky Survey \citep[2MASS;][]{skrutskie2006}, and the Panoramic Survey Telescope and Rapid Response System \citep[Pan-STARRS;][]{kaiser2002}.
The dust distribution was reconstructed using variational inference and Gaussian process regression \citep[MGVI,][]{knollmuller2018,leike2019}.
This method is focused on modeling the dust distribution in the vicinity of the Sun, in contrast with other 3D mapping efforts aimed at reconstructing the distribution of dust in our Galaxy on large scales to study the structure of our Galaxy, such as its spiral arms \citep{lallement2018,chen2019,green2019}.

The final data product from \cite{leike2020} is a reconstruction of the G-band opacity per parsec, $s_{x}$\,$\equiv$\,$(\Delta\tau_{\rm G})/(\Delta L/{\rm pc})$, within 400\,pc from the Sun with a resolution of up to $1$\,pc along the line of sight (LOS).
Following the procedure described in \cite{bialy2021}, we converted $s_{x}$ into the total volume density of hydrogen nuclei $n_{\rm H}$.
We assumed a G-band extinction per hydrogen nuclei column density $A_{\rm G}/N_{H}$\,$=$\,$4$\,$\times$\,$10^{-22}$\,mag\,cm$^{2}$ \citep[][]{draine2011}, which leads to 
\begin{equation}
n = 880s_{x}\,{\rm cm}^{-3}.
\end{equation}

We calculated $n_{\rm H}$ along the LOS by placing an observer in the position of the Sun and tracing radial rays for which we computed $s_{x}$ in 1-pc steps up to 588\,pc.
We registered the radial $n_{\rm H}$ distribution in 588 {\tt HEALPix} \citep{gorski2005} spheres with tessellation $N_{\rm side}$\,$=$\,$512$, which corresponds to pixel angular sizes $\theta_{\rm pix}$\,$=$\,6\parcm87.
We extracted the observed dust distribution toward Taurus by making Cartesian projections that match the 20\arcsec\ grid in the $^{12}$CO and $^{13}$CO data.
An example of the resulting maps is shown in  Fig.~\ref{fig:HIandCOand3Ddust_maps}.

\subsubsection{Dust column density}

We also used the dust optical depth at 353\,GHz ($\tau_{353}$) as a proxy for the \juan{total hydrogen nuclei}  column density ($N_{\rm H}$).
The $\tau_{353}$ map was derived from the all-sky {\it Planck} intensity observations at 353, 545, and 857\,GHz, and the IRAS observations at 100\,$\mu$m, which were fitted using a modified black body spectrum \citep{planck2013-p06b}.
Other parameters obtained from this fit are the temperature and the spectral index of the dust opacity. 
The angular resolution of the resulting $\tau_{353}$ is 5\arcmin\ FWHM.

We scaled from $\tau_{353}$ to $N_{\rm H}$ using the Galactic extinction measurements toward quasars \citep{planck2013-p06b}, which lead to:
\begin{equation}\label{eq:nh}
\tau_{353}/N_{H} = 1.2 \times 10^{-26}\,{\rm cm}^{2}.
\end{equation}
Variations in dust opacity are present even in the diffuse ISM and the opacity increases systematically by a factor of two from the diffuse to the denser ISM \citep[see, for example,][]{planck2011-7.12,martin2012}. 
However, our results do not critically depend on this calibration.

\section{Methods}\label{sec:methods}

\subsection{Comparing spectral emission and 3D dust cubes}

We correlated the H{\sc i} and CO emission across velocity channels and the 3D dust cube using the histogram of oriented gradients (HOG) method introduced in \cite{soler2019a}.
This method is based on characterizing the emission distribution using the orientation of its gradients.
For a pair of intensity maps $I_{l}^{\rm A}$ and $I_{m}^{\rm B}$ in \juan{the emission position-position-velocity (PPV)} cubes A and B and velocity channels $l$ and $m$ (or a distance slice in the 3D density cube), the relative orientation between their gradients is:
\begin{equation}\label{eq:phi}
\theta_{ij, lm}=\arctan\left(\frac{(\nabla I^{\rm A}_{ij,l} \times \nabla I^{\rm B}_{ij,m})\cdot\hat{z}}{\nabla I^{\rm A}_{ij,l} \cdot \nabla I^{\rm B}_{ij,m}}\right),
\end{equation}
where the $i$ and $j$ indexes run over the pixels in the sky coordinates, which in our case are right ascension (RA) and declination (DEC), and $\nabla$ is the differential operator that corresponds to the gradient.

Equation~\eqref{eq:phi} implies that the relative orientation angles are in the range $[-\pi/2,\pi/2)$, thus accounting for the orientation of the gradients and not their direction.
The values of $\theta_{ij,lm}$ are only meaningful in regions where both $|\nabla I^{\rm A}_{ij,l}|$ and $|\nabla I^{\rm B}_{ij,m}|$ are greater than zero or above thresholds that are estimated according to the noise properties of the each PPV cube (3D density cube).

We explicitly compute the gradients using Gaussian derivatives by applying the multidimensional Gaussian filter routines in the {\tt filters} package of {\tt Scipy}.
The Gaussian derivatives result from the image's convolution with the spatial derivative of a two-dimensional Gaussian function.
The width of the Gaussian determines the area of the vicinity over which the gradient is calculated.
Varying the width of the Gaussian kernel enables the sampling of different scales and reduces the effect of noise in the pixels \citep[see,][and references therein]{soler2013}.

We synthesize the morphological correlation contained in $\theta_{ij, lm}$ by summing over the spatial coordinates, indexes $i$ and $j$, using the projected Rayleigh statistic \citep[][]{jow2018}, which is defined as
\begin{equation}\label{eq:prs}
V_{lm} = \frac{\sum_{ij}w_{ij,lm}\cos(2\theta_{ij,lm})}{\left(\sum_{ij}w_{ij,lm}/2\right)^{1/2}}.
\end{equation}
The projected Rayleigh statistic ($V$) is a test of non-uniformity in a distribution of angles for the specific orientation of interest $\theta_{0}$\,$=$\,$0\deg$ and $90\deg$, such that $V$\,$>$\,0 or $V$\,$<$\,0 correspond to clustering around those two particular angles \citep{durandANDgreenwood1958}.
Values of $V$\,$>$\,0, which imply that the gradients are mostly parallel, quantify the significance of the morphological similarity between the two maps $I_{l}^{\rm A}$ and $I_{m}^{\rm B}$.
Values of $V$\,$<$\,0, which imply that the gradients are mostly perpendicular, are relevant in magnetic field studies \citep[see, for example,][]{heyer2020}.

The null hypothesis implied in $V$ is that the angle distribution is uniform.
In the particular case of independent and uniformly distributed angles and for a large number of samples, values of $V$\,$\approx$\,$1.64$ and $2.57$ correspond to the rejection of the null hypothesis with a probability of 5\% and 0.5\%, respectively \citep{batschelet1972}.
Thus, a value of $V$\,$\approx$\,2.87 is roughly equivalent to a 3$\sigma$ confidence interval.
Similarly to the chi-square test probabilities, $V$ and its corresponding null hypothesis rejection probability are reported in the classical circular statistics literature as tables of ``critical values'', for example, in \cite{batschelet1972}, or computed in the circular statistics packages, such as {\tt circstats} in {\tt astropy} \citep{astropy2018}.

We accounted for the spatial correlations introduced by the \juan{scale of the derivative kernel} by choosing the statistical weights $w_{ij,lm}$\,$=$\,$(\delta x/\Delta)^{2}$, where $\delta x$ is the pixel size and $\Delta$ is the \juan{derivative kernel's FWHM}.
For pixels where the norm of the gradient is negligible or can be confused with the signal produced by noise, we \juan{set} $w_{ij,lm}$\,$=$\,0.
If all the gradients in an image pair are negligible, Eq.~\ref{eq:prs} has an indeterminate form that is treated numerically as Not a Number ({\tt NaN}).

\juan{The size of the derivative kernel sets a common spatial scale for comparing two maps.
Thus, we do not smooth the input data to the same angular resolution for the HOG computation with derivative kernels larger than the angular resolution of the beam.
However, smoothing the input data to a coarser angular resolution increases the signal-to-noise ratio, $(S/N)$, thus increasing the number of gradients in the HOG computation.
In the main body of this paper, we report the results obtained using the input data in its native angular resolution and show the results of pre-smoothing in Appendix~\ref{appendix:smoothing}.
This selection does not significantly change the conclusions of our study.}

%
%

\juan{We compared the morphological correlation evaluated by $V$ with the correlation in the amount of emission or density by calculating} the Pearson correlation coefficient, which is defined as 
\begin{equation}\label{eq:pearsonr}
(r_{\rm Pearson})_{lm} = \frac{\sum_{ij}\left(I^{\rm A}_{ij,l}-\left<I^{\rm A}_{l}\right>\right)\left(I^{\rm B}_{ij,l}-\left<I^{\rm B}_{l}\right>\right)}{\sqrt{\sum_{ij}\left(I^{\rm A}_{ij,l}-\left<I^{\rm A}_{l}\right>\right)^{2}}\sqrt{\sum_{ij}\left(I^{\rm B}_{ij,l}-\left<I^{\rm B}_{l}\right>\right)^{2}}},
\end{equation}
where $\left<I^{\rm A}_{l}\right>$ and $\left<I^{\rm B}_{l}\right>$ are the weighted mean values of the maps introduced in Eq.~\eqref{eq:phi}.
The Pearson correlation coefficient is a normalized measurement of the covariance and is a measure of linear correlation between the two tracers.
In contrast to $V$, it is sensitive to the amount of emission rather than the spatial distribution of the tracers across the map.

\section{Results}\label{sec:results}

\subsection{Dust in 2D and 3D}

\begin{figure}[ht!]
\centerline{\includegraphics[width=0.49\textwidth,angle=0,origin=c]{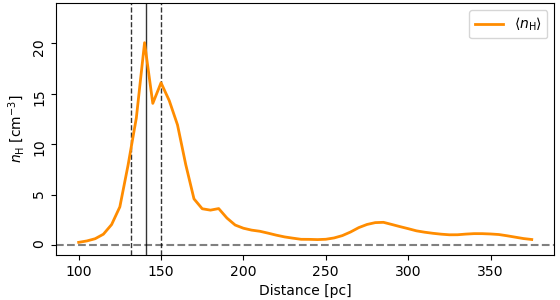}}
\centerline{\includegraphics[width=0.49\textwidth,angle=0,origin=c]{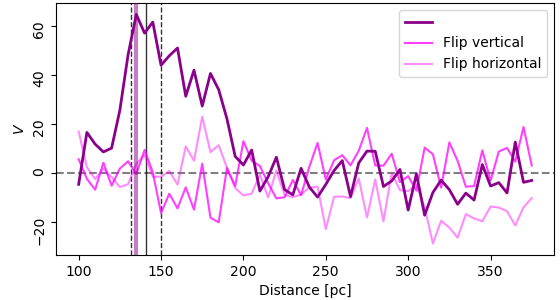}}
\centerline{\includegraphics[width=0.49\textwidth,angle=0,origin=c]{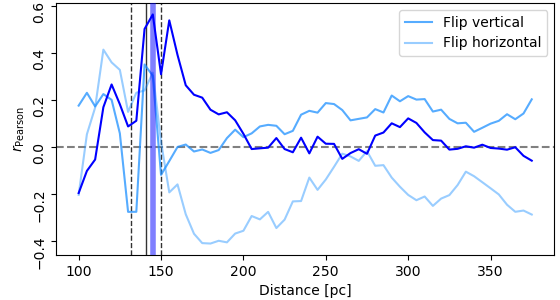}}
\caption{
{\it Top}. Mean volume density of hydrogen nuclei ($n_{\rm H}$) derived from the 3D dust reconstruction toward the region shown in Fig.~\ref{fig:HIandCOand3Ddust_maps}.
{\it Middle}. Projected Rayleigh statistic ($V$, Eq.~\ref{eq:prs}) for the integrated dust column density derived from the {\it Planck} observations ($N_{H}$) and the distribution of dust across distance slices in the 3D dust reconstruction.
Values of $V$\,$\approx$\,0 correspond to a random orientation between the gradients of the two tracers, thus indicating low morphological correlation.
Values of $V$\,$>$\,2.87 correspond to the mostly parallel gradient, thus indicating a significant morphological correlation.
{\it Bottom}. Pearson correlation coefficient ($r_{\rm Pearson}$) between $N_{H}$ and the distribution of dust across distance slices.
The \juan{solid} colored vertical line indicates the distance channel with the maximum $V$.
The faint lines correspond to the chance correlation tests performed by flipping the maps with respect to each other in the vertical and horizontal directions.
\juan{The black vertical lines indicate the average distance to the Taurus MC and its corresponding confidence interval, as reported in \cite{zucker2020}}.
}
\label{fig:HOGdustANDdust3D}
\end{figure}

We applied the HOG method to the integrated dust column density derived from the {\it Planck} observations ($N_{\rm H}$) and the $n_{\rm H}$ distribution across distance bins using a $\Delta$\,$=$\,\juan{30}\arcmin\ FWHM derivative kernel, \juan{which corresponds to a scale of approximately 1.2\,pc at the mean distance to Taurus.}
To evaluate the impact of random correlation, we also present the results obtained when flipping the $N_{\rm H}$ map vertically and horizontally.
The resulting $V$ and $r_{\rm Pearson}$ are presented in Fig.~\ref{fig:HOGdustANDdust3D}.

The morphological correlation between these two data sets, as quantified by $V$, presents maximum values around the distances usually assigned to the Taurus MC.
Most of the dust distribution roughly beyond 200\,pc shows very small $V$ values.
The flipped $N_{\rm H}$ maps produce relatively low values of $V$ with respect to the peak of the values in the original maps, thus suggesting that the effect of chance correlation is negligible.

The concentration of high $V$ around the distance of the MC indicates that most of the integrated $N_{\rm H}$ distribution results from the contributions from the dust in the Taurus MC with negligible input from the material in the background, \juan{beyond 200\,pc}.
The \juan{$V$ distribution across distances suggests} that the morphological matching obtained with the HOG method provides \juan{an alternative} definition of MC distance.
This ``morphological distance''\juan{, understood as the range of line of sight distances that contribute the most to the observed morphology}, is complementary to \juan{the distance defined} using reddening and distances to stars \citep[see, for example,][]{green2014,schlafly2014,zucker2020}.

The Pearson correlation coefficient ($r_{\rm Pearson}$) also shows maximum values for the distances around the MC.
However, the results of flipping the $N_{\rm H}$ map indicate that $r_{\rm Pearson}$ presents large values produced by chance correlation.
We interpret these results as indicating that $V$ is less sensitive to chance correlation than the linear correlation quantified by $r_{\rm Pearson}$.

\juan{The increase in the values of $r_{\rm Pearson}$ roughly matches with the increase in the mean density along the line of sight.
In contrast, $V$ appears to be dominated by the density peak around the distance of Taurus.
This indicates that the Taurus MC is the dominant feature in the $N_{\rm H}$ map, although it is not the only density structure found along the line of sight.}

\begin{figure*}[ht!]
\centerline{
\includegraphics[width=0.49\textwidth,angle=0,origin=c]{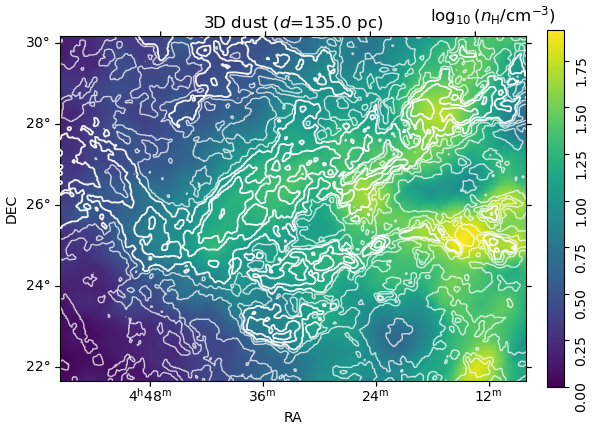}
\includegraphics[width=0.49\textwidth,angle=0,origin=c]{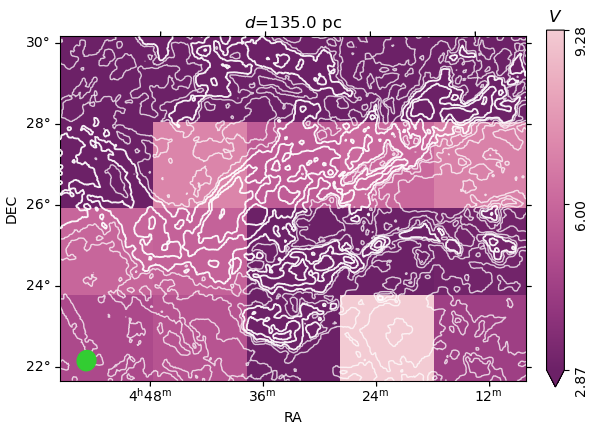}
}
\caption{
{\it Left}. Distribution of the hydrogen volume density ($n_{\rm H}$), \juan{derived from the \cite{leike2020} 3D dust reconstruction, for the distance channel that shows the highest correlation with the gas column density ($N_{\rm H}$) derived from the {\it Planck} observations, shown by the contours.
The white contours correspond to $N_{\rm H}$\,$=$\,1, 2, 3, 4, 5, 8, and 12.5\,$\times$\,$10^{21}$\,cm$^{-2}$.}
{\it Right}. Distribution of the projected Rayleigh statistic ($V$, Eq.~\ref{eq:prs}) in \juan{5\,$\times$\,4} blocks across the map.
\juan{The minimum of the color bar is set to 2.87, which is roughly equivalent to a 3$\sigma$ significance for this indicator of the correlation between the two images.
}
The green disk indicates the \juan{derivative kernel} size $\Delta$\,$=$\,30\arcmin.
}
\label{fig:HOGdustANDdust3Dmaps}
\end{figure*}

\juan{Figure~\ref{fig:HOGdustANDdust3Dmaps} shows the $n_{\rm H}$ distribution in the slice with the highest $V$ and the $V$ distribution in 5\,$\times$\,4 blocks, each roughly $4\Delta$ in side, for that distance channel.
We found a significant correlation between $n_{\rm H}$ and $N_{\rm H}$ toward the regions around the dark clouds B220/L1527 and B211/B213/L1495.
There is also significant $V$ toward mostly diffuse regions, for example, in the southeast portion of the map.
The highest $V$ is found toward the shell-like feature in the southwest portion of the map, which is not prominent in the near-infrared extinction maps of Taurus \citep[see, for example,][]{lombardi2010}.}

\subsection{Gas and 3D dust}

\begin{figure*}[ht!]
\centerline{
\includegraphics[width=0.33\textwidth,angle=0,origin=c]{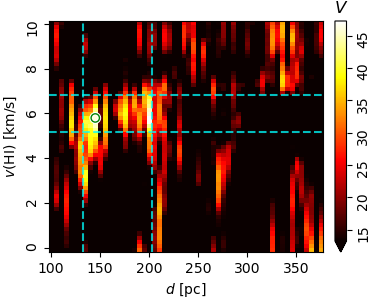}
\includegraphics[width=0.33\textwidth,angle=0,origin=c]{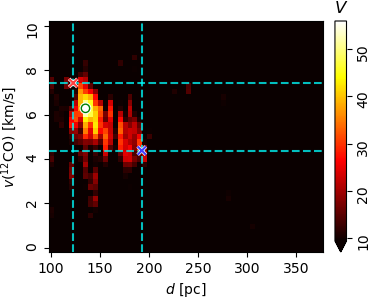}
\includegraphics[width=0.33\textwidth,angle=0,origin=c]{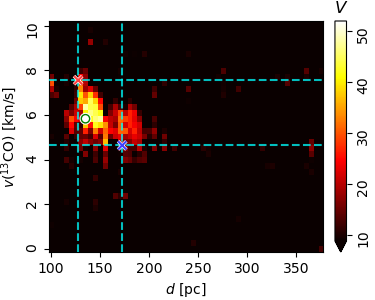}
}
\caption{Correlation between the distribution of hydrogen volume density ($n_{\rm H}$) derived from the \cite{leike2020} 3D dust reconstruction and the H{\sc i}, $^{12}$CO, and $^{13}$CO emission across distances and radial velocities as quantified by the projected Rayleigh statistic ($V$), as indicated in the color bars to the right of each panel.
\juan{The lower limit of each color scale is set to the standard deviation of $V$ across the whole distance and velocity range, $\varsigma_{V}$. 
The dashed vertical and horizontal lines indicate the minimum and maximum $d$ and $v_{\rm LSR}$ where $V$ is larger than 3$\varsigma_{V}$.}
The circle indicates the distance-radial velocity pair with the highest $V$.
The red and blue crosses in the center and left panels indicate the highest radial velocity and nearest distance and the lowest radial velocity and farthest distances with significant $V$, respectively.
}
\label{fig:CollapseTestVplane}
\end{figure*}

We also applied the HOG method to the H{\sc i}, $^{12}$CO, and $^{13}$CO emission and the dust distribution across distance bins using a $\Delta$\,$=$\,\juan{30}\arcmin\ FWHM derivative kernel.
The distribution of $V$ across distance slices and radial velocity channels is presented in Fig.~\ref{fig:CollapseTestVplane}.
 The maximum values of $V$ indicate a large correlation in the distribution of the dust and gas tracers.
 However, there are clear differences between the results obtained for the H{\sc i} and the two CO isotopologues.
 
\subsubsection{H{\sc i} and 3D dust}

The distribution of $V$ for the H{\sc i} emission and the 3D dust, shown in the left-hand side panel of Fig.~\ref{fig:CollapseTestVplane}, indicates a large spread of relatively high values for a broad range of velocity channels and distances.
The highest $V$ values are found in the range 130\,$\lesssim$\,$d$\,$\lesssim$\,160\,pc and 3\,$\lesssim$\,$v_{\rm LSR}$\,$\lesssim$\,7\,km\,s$^{-1}$.
Yet, we also found high $V$ for a few velocity channels around 5.7\,km\,s$^{-1}$ at $d$\,$\approx$\,200\,pc.

The multiplicity of distances with high $V$ for a particular $v_{\rm LSR}$ implies that the H{\sc i} emission in a velocity channel results from contributions of gas parcels separated by tens of parsecs.
This result is not surprising given the ubiquity of H{\sc i} and the relatively broad linewidths; around 2\,km\,s$^{-1}$ and 10\,km\,s$^{-1}$ for the cold and warm H{\sc i} phases \citep[see, for example,][]{kalberla2009}.
The overlap is particularly acute for $v_{\rm LSR}$\,$\approx$\,5.7\,km\,s$^{-1}$, where there are high $V$ values at $d$\,$\approx$\,125, 140, and 200\,pc.

\begin{figure*}[ht!]
\centerline{
\includegraphics[width=0.49\textwidth,angle=0,origin=c]{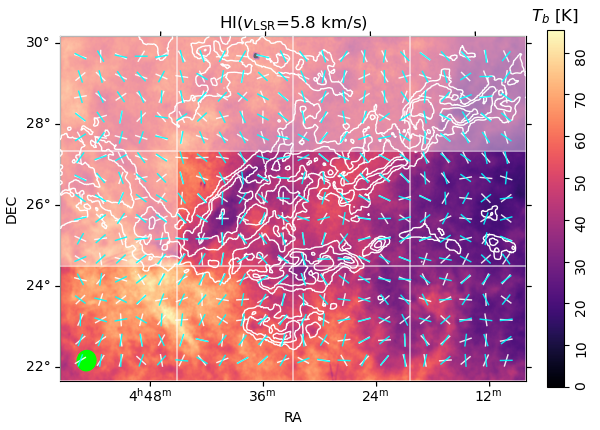}
\includegraphics[width=0.49\textwidth,angle=0,origin=c]{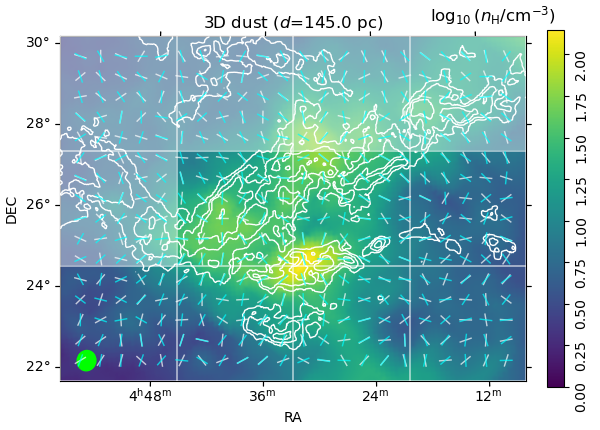}
}
\centerline{
\includegraphics[width=0.49\textwidth,angle=0,origin=c]{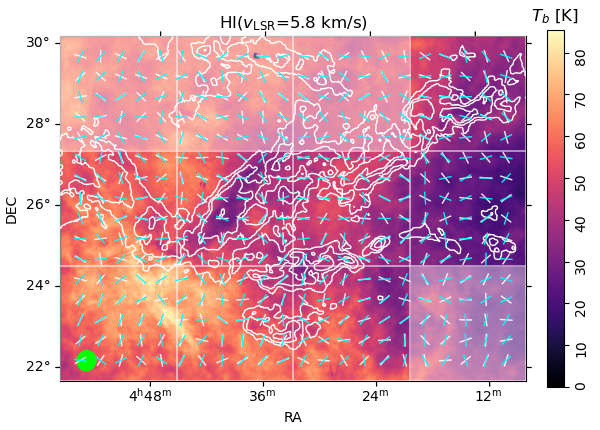}
\includegraphics[width=0.49\textwidth,angle=0,origin=c]{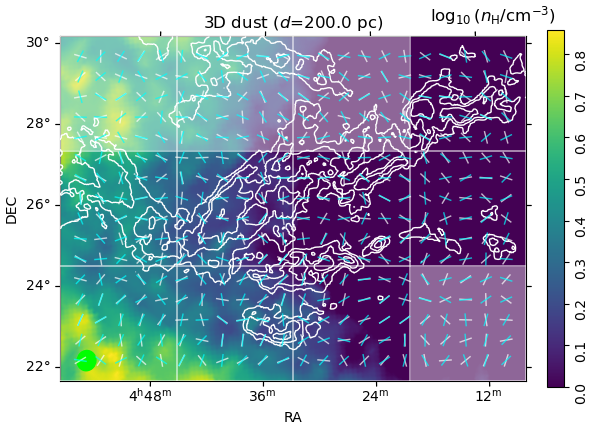}
}
\caption{Atomic hydrogen (H{\sc i}) emission and distribution of the hydrogen volume density ($n_{\rm H}$) derived from the \cite{leike2020} 3D dust reconstruction for two radial velocity and distance channels with the highest $V$, as \juan{indicated} in Fig.~\ref{fig:CollapseTestVplane}.
\juan{The pseudovectors indicate the orientation of the $T_{b}$ and $n_{\rm H}$ gradients, shown in white and cyan colors respectively.}
The translucent mask marks the blocks where $V$\,$<$\,2.87, thus indicating areas of negligible morphological correlation.
}
\label{fig:maxVdustandHI}
\end{figure*}

The top panel in Figure~\ref{fig:maxVdustandHI} shows the $n_{\rm H}$ distribution and H{\sc i} emission for the pair of distance and radial velocity channels with the highest $V$\juan{, indicated by the green circle in the corresponding panel of Fig.~\ref{fig:CollapseTestVplane}}.
The distance with the highest $V$ is \juan{found around $v_{\rm LSR}$\,$\approx$\,5.8\,km\,s$^{-1}$ and $d$\,$\approx$\,145\,pc, which indicates that there is a significant correlation between dust in and around the MC and the H{\sc i} emission.}
We found, however, that the highest similarity between the tracers comes from regions within Taurus where the increase in $n_{\rm H}$ is associated with a decrease in the H{\sc i}.
Such anti-correlation is expected if we consider that the H{\sc i} associated with the MC is in the cold phase that is seen as a shadow against the warm phase background, which is known in the literature as H{\sc i} self-absorption  \citep[HISA;][]{heeschen1955,gibson2000,wang2020hisa}.

\juan{We also found significantly high $V$ for $v_{\rm LSR}$\,$\approx$\,5.8\,km\,s$^{-1}$ and $d$\,$\approx$\,200\,pc.
The distribution of dust around that distance, shown in the bottom panel of Figure~\ref{fig:maxVdustandHI}, corresponds to the H{\sc i} bright region toward the left-hand side of the map, which most likely corresponds to the background against which the HISA around the B220/L1527 region in Taurus is visible.
However, the bright H{\sc i} filament seen over the more extended emission toward the lower left portion of the map does not have a counterpart in the dust distribution at that distance.}

\subsubsection{CO and 3D dust}

\begin{figure*}[ht!]
\centerline{
\includegraphics[width=0.49\textwidth,angle=0,origin=c]{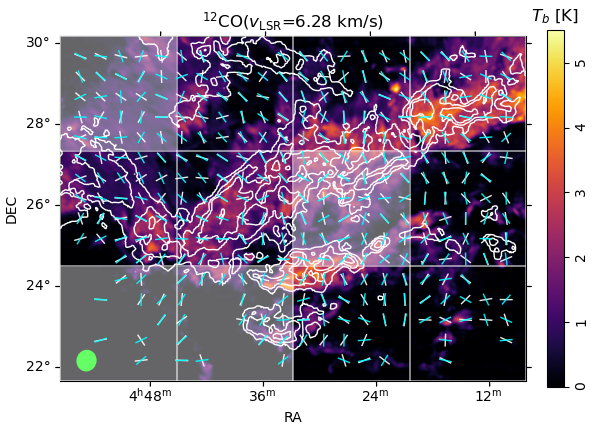}
\includegraphics[width=0.49\textwidth,angle=0,origin=c]{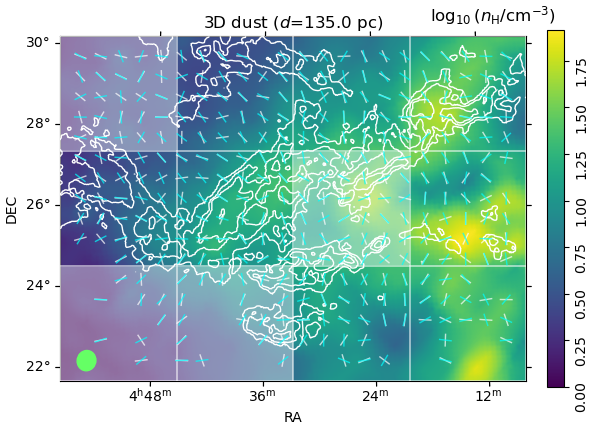}
}
\caption{Carbon monoxide ($^{12}$CO) emission and distribution of the hydrogen volume density ($n_{\rm H}$) derived from the \cite{leike2020} 3D dust reconstruction for the radial velocity and distance channels with the highest $V$, as \juan{indicated} in Fig.~\ref{fig:CollapseTestVplane}.
\juan{The pseudovectors indicate the orientation of the $T_{b}$ and $n_{\rm H}$ gradients, shown in white and cyan colors respectively.}
The translucent mask marks the blocks where $V$\,$<$\,2.87, thus indicating areas of negligible morphological correlation.
}
\label{fig:maxVdustand12CO}
\end{figure*}

The distribution of $V$ for the $^{12}$CO and $^{13}$CO emission and the 3D dust, shown on the center and right panels of Fig.~\ref{fig:CollapseTestVplane}, indicates a concentration of high values in the range 130\,$\lesssim$\,$d$\,$\lesssim$\,160\,pc and 4\,$\lesssim$\,$v_{\rm}$\,$\lesssim$\,7\,km\,s$^{-1}$.
In the case of $^{13}$CO, the highest $V$ appear closely concentrated around $d$\,$\approx$\,140\,pc and $v_{\rm LSR}$\,$\approx$\,6.2\,km\,s$^{-1}$.
In the case of $^{12}$CO, the highest $V$ values appear more spread out in velocity and distance.
They also show a negative slope distribution in the distance-radial velocity plane that links the highest distances to the lowest velocities and the lowest distances to the highest velocities.

Fig.~\ref{fig:maxVdustand12CO} shows the $n_{\rm H}$ distribution and $^{12}$CO emission for the distance and radial-velocity channels with the highest $V$.
The similarity in the distribution of both tracers confirms the results of the HOG. 
Most of the correlation comes from the material within the 5\,$\times$\,$10^{21}$\,cm$^{-2}$ ($A_{V}$\,$\approx$\,5) contour, which is around the extinction level usually used to define the Taurus MC in extinction and dust emission observations \citep[see, for example,][]{cambresy1999,planck2015-XXXV}.
There are, however, small portions of the cloud for which the $n_{\rm H}$ distribution has no counterpart in the $^{12}$CO emission in the studied velocity range, particularly toward the lower right corner of the map.

\begin{figure*}[ht!]
\centerline{
\includegraphics[width=0.99\textwidth,angle=0,origin=c]{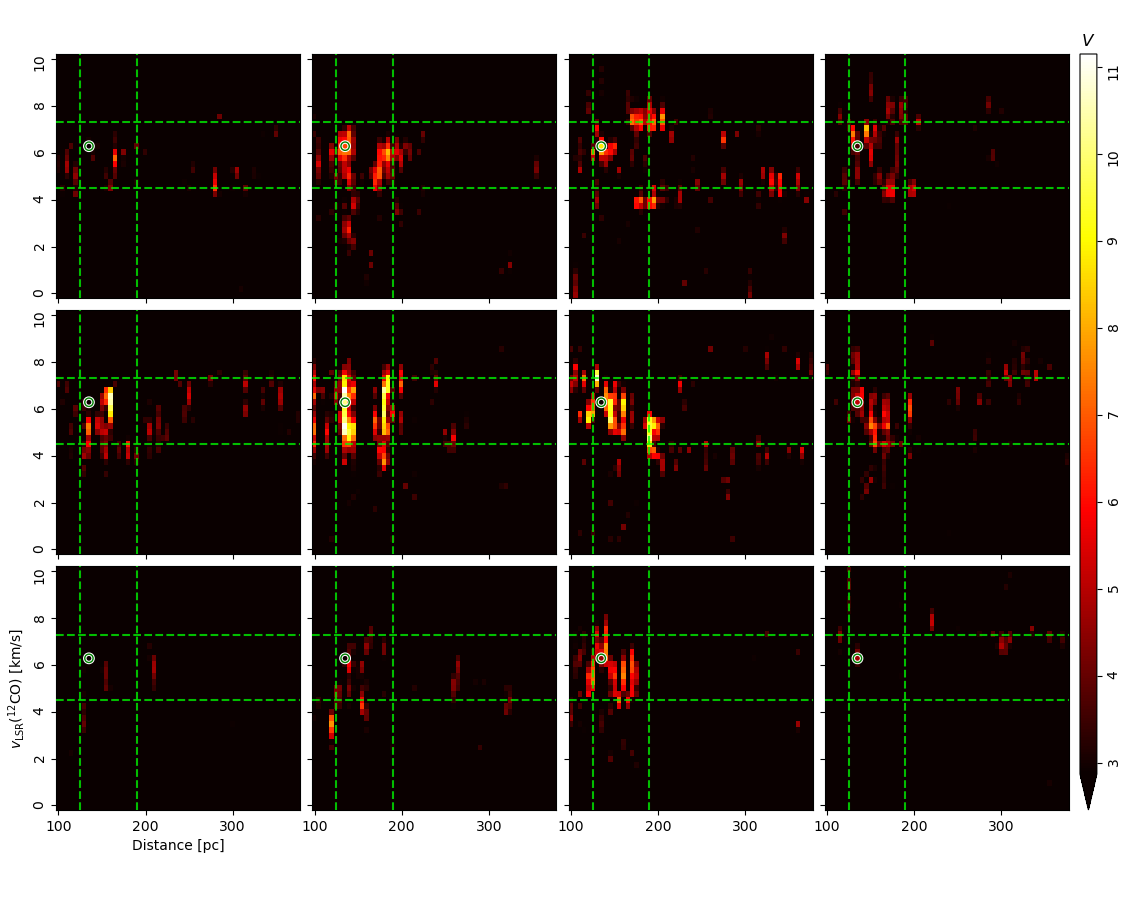}
}
\caption{Correlation between the distribution of the hydrogen volume density ($n_{\rm H}$) derived from the \cite{leike2020} 3D dust reconstruction and $^{12}$CO emission across distances and radial velocities as quantified by the projected Rayleigh statistic ($V$) for the 4\,$\times$\,3 regions shown in Fig.~\ref{fig:maxVdustand12CO}.
\juan{The vertical and horizontal dashed lines correspond to the distances and radial velocity ranges defined in Fig.~\ref{fig:CollapseTestVplane}.
The circle marks the pair of distance and velocity channels with the highest global $V$, which are presented in Fig.~\ref{fig:maxVdustand12CO}.}
}
\label{fig:blockaverageVplane12CO}
\end{figure*}

We further explored the negative-slope feature in the center panel of Fig.~\ref{fig:CollapseTestVplane} by dividing the area of the Taurus MC in 4\,$\times$\,3 blocks and calculating the $V$ distribution across distances and radial velocities.
The results toward each block, presented in Fig.~\ref{fig:blockaverageVplane12CO}, indicate that the signal in Fig.~\ref{fig:CollapseTestVplane} is also found in the upper right areas of the Taurus MC, with the largest significance found toward the center of the region.
The block with the highest column density, second from left to right in the central row of Fig.~\ref{fig:blockaverageVplane12CO}, shows a discontinuity most likely related to the high extinction limiting the 3D dust reconstruction toward that portion of the cloud.
Insofar as the division in blocks does not elucidate the dynamical behavior of each portion of the cloud and is limited by the limited number of independent gradients in each block, we focus on the global trend rather than look for an explanation of the trend for each MC portion.

The global HOG results presented in Fig.~\ref{fig:CollapseTestVplane} indicate that the material at the highest distances is related to the lowest velocities and that material at lower distances is linked to the highest velocities, but it is not limited to these two extremes.
This high correlation identified with the HOG method is distributed across the distances and radial velocities between roughly 125 and 175\,pc and 4 to 7\,km\,s$^{-1}$.
This distribution implies that the MC's most distant and closest portions are converging along the line of sight.

The distinctive pattern in $V$ linking the near side of Taurus to its highest $^{12}$CO radial velocities and vice versa is slightly distinguishable in the results obtained with $^{13}$CO, shown in the right panel of Fig.~\ref{fig:CollapseTestVplane}.
However, its lower significance limits additional interpretation, as further illustrated in Appendix~\ref{appendix:blocks}.

\section{Discussion}\label{sec:discussion}

\begin{figure*}[ht!]
\centerline{
\includegraphics[width=0.49\textwidth,angle=0,origin=c]{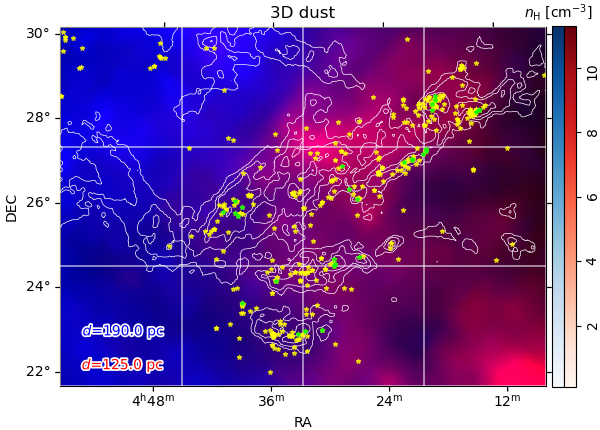}
\includegraphics[width=0.49\textwidth,angle=0,origin=c]{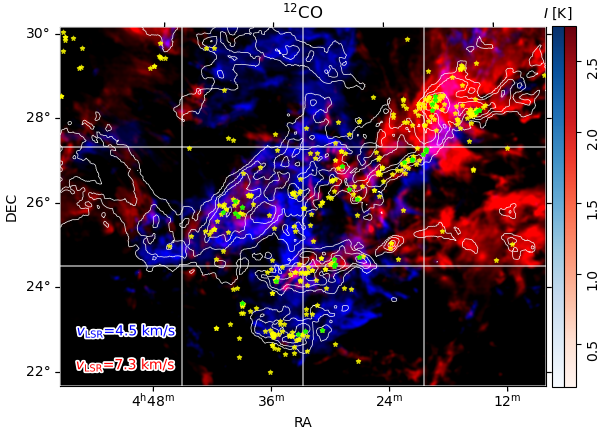}
}
\caption{Distribution of the hydrogen volume density ($n_{\rm H}$) derived from the \cite{leike2020} 3D dust reconstruction (left) and $^{12}$CO emission (right) for the minimum and maximum distance and radial velocity channels with significant correlation, \juan{ identified by the red and blue crosses} in Fig.~\ref{fig:CollapseTestVplane}.
\juan{The maps shown in a blue colormap correspond to the farthest distance and the lowest velocity channels with $V$\,$>$\,3$\varsigma_{V}$, where $\varsigma_V$ is the standard deviation of $V$ across the whole distance and velocity range.}
\juan{The maps shown in a red colormap correspond to the nearest distance and the highest velocity channels with $V$\,$>$\,3$\varsigma_{V}$.}
The yellow and green stars indicate the positions of the young (age\,$<$\,10\,Myr) stellar objects presented in \cite{galli2019} and young stellar clusters in \cite{luhman2022}, respectively.
}
\label{fig:3Ddustand13COrgb}
\end{figure*}

\subsection{Morphology as an indicator of distance}

Figure~\ref{fig:HOGdustANDdust3D} illustrates the agreement between the estimated distance to the Taurus MC and the distances at which the HOG produced the highest correlation, quantified by $V$.
The HOG produces high $V$ when the gradients of two images are mostly parallel. 
Thus, the finding of high $V$ at the distance of the MC indicates that the dust parcel at $d$\,$=$\'140\,pc not only has most of the dust along the line of sight, but it is also responsible for most of the features observed in the integrated dust map.
This observation is essential for the comparison between the 2D MCs in the literature, which are traditionally in the integrated dust maps, and the novel reconstructions of 3D dust distribution \citep[see, for example,][]{zucker2021,rezaei2022}.

The tests of our method applied to the 3D maps of dust reddening from \cite{green2019} emphasize the importance of the 3D dust reconstruction for the HOG studies, as detailed in Appendix~\ref{appendix:extinctionmaps}.
The 3D dust maps have the advantage of accounting for the spatial correlations between independent lines of sight, producing a smoother distribution.
Moreover, the 3D maps of dust reddening tend to overestimate the cloud distance obtained with the HOG and saturate for distances beyond the dominant feature in the map, as discussed in Appendix~\ref{appendix:extinctionmaps}.

We found no significant signal in $V$ at $d$\,$\approx$\,300\,pc, corresponding to the backside of the Per-Tau shell identified in \cite{bialy2021}.
This lack of correlation indicates that this potential background does not significantly contribute to the integrated dust emission and emphasizes the importance of the 3D dust maps to elucidate this kind of structure surrounding MCs. 
Being at a relatively high Galactic latitude ($b$\,$\approx$\,$-$14\pdeg5), the integrated dust emission toward Taurus is unlikely to be affected by background contamination.
However, this may be critical for clouds at lower $b$, where the morphological correlation quantified by the HOG can identify what parcel of dust along the LOS is responsible for the integrated emission and what portions of an MC are dominated by different dust parcels along the LOS.

\subsection{Dust revealing the presence of cold H{\sc i}}

The anti-correlation between H{\sc i} emission and $n_{\rm H}$ illustrated in Fig.~\ref{fig:maxVdustandHI} marks the potential location of HISA in Taurus.
Observations of H{\sc i} in dark clouds in Taurus have previously revealed the presence of H{\sc i} narrow self-absorption (HINSA) features, suggesting that a significant fraction of the H{\sc i} in Taurus is located in the cold, well-shielded portions of MCs and is mixed with the molecular gas \citep{li2003,li2005}.

\cite{heiner2013} used the H{\sc i} spectra to identify HISA and estimate hydrogen volume densities of up to 430\,cm$^{-3}$ within Taurus.
This value is at least a factor of two larger than the maximum $n_{\rm H}$ derived from the 3D dust across the MC.
This discrepancy can be explained by the limitation of the 3D dust to reconstruct the density toward regions of high extinction.

Given the lack of existing observations of H{\sc i} absorption around Taurus, just one within 5\deg\ of the center of the region \citep{nguyen2019}, the correlation between the dust and the HISA is a promising path to disentangle the amount of cold H{\sc i} in and around MCs.
We will focus on this topic in the subsequent paper in this series.

\subsection{Dynamics of the Taurus MC}\label{sec:discussionDynamics}

The negative slope feature found in the correlation between $n_{\rm H}$ and $^{12}$CO emission, shown in the central panel of Fig.~\ref{fig:CollapseTestVplane}, indicates that nearby material in Taurus is moving toward the central velocity of the cloud and more distant material is also moving toward it.
This motion is consistent with two scenarios.
One possibility is rotation on MC scales.
The other possibility is the MC is contracting, either by external compression or global collapse.


Taking the cloud as a whole, our results are consistent with a 2.5\,km\,s$^{-1}$ radial velocity difference more or less aligned with the Galactic plane across roughly 25\,pc, that is, a velocity gradient of roughly 0.1\,km\,s$^{-1}$\,pc$^{-1}$.
\cite{meidt2018} have shown that the gas kinematics on the scale of individual molecular clouds are not entirely dominated by self-gravity but also track a component that originates from orbital motion in the potential of the host galaxy. 
\cite{pascucci2015} presents observations of the narrow Na and K absorption lines toward 40 T-Tauri stars in the Taurus and found a velocity gradient along the length of the MCs, which the authors suggest is consistent with differential galactic rotation \citep{imara2011}.

\cite{galli2019} studied the parallax distances and proper motions from {\it Gaia}'s Data Release 2 (DR2) and the astrometry obtained with long baseline interferometry (VLBI) for 415 stars towards the Taurus MC.
Using the divergence and rotational of stellar proper motions, the authors concluded that the expansion or contraction motions traced by this sample are negligible.
They also reported a global rotational velocity of around 1.5\,km\,s$^{-1}$.
However, the results of the HOG indicate a different scenario for the dust and the gas.
 
Figure~\ref{fig:3Ddustand13COrgb} shows the $n_{H}$ distribution and the $^{12}$CO emission for the extremes of the negative-slope feature identified in Fig.~\ref{fig:CollapseTestVplane}
The colors indicate what can be interpreted as a large-scale velocity gradient between $v_{\rm LSR}$\,$=$\,4.25\,km\,s$^{-1}$ and 7.05\,km\,s$^{-1}$ across Taurus.
However, the $^{12}$CO emission toward the portion of the MC that shows the HOG signature of compression, as identified in Fig.~\ref{fig:blockaverageVplane12CO}, shows a less clear left-right symmetry.
The $n_{\rm H}$ for the two distances associated with the radial velocities shows a mixture of two components, and it is mainly dominated by the nearby one at $d$\,$=$\,125\,pc.

\begin{figure}[ht!]
\centerline{
\includegraphics[width=0.49\textwidth,angle=0,origin=c]{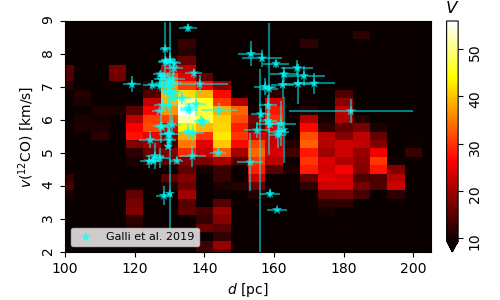}
}
\centerline{
\includegraphics[width=0.49\textwidth,angle=0,origin=c]{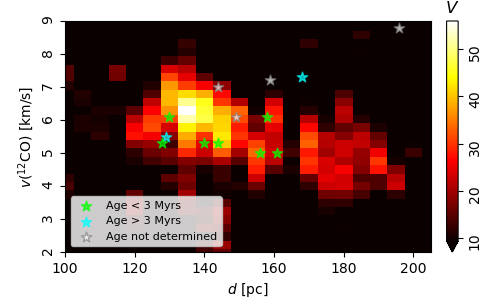}
}
\caption{Same as the middle panel of Fig.~\ref{fig:CollapseTestVplane}, but for a more limited range of distances and radial velocities and including the radial velocities and distances for the young (age\,$<$\,10\,Myr) stellar objects identified as cluster members in \cite{galli2019} and young stellar clusters in \cite{luhman2022}.}
\label{fig:prs3Ddustand13COwithYoungClusters}
\end{figure}

The comparison between the HOG results and the stellar distances and radial velocities in the \cite{galli2019} sample, shown on the top panel of Fig.~\ref{fig:prs3Ddustand13COwithYoungClusters}, suggest that the stars follow the trend in the gas and dust, albeit with a large scatter.
The scatter in the stellar data is explained by the 2.7\,km\,s$^{-1}$ three-dimensional velocity dispersion reported in \cite{galli2019}, which is close to twice the rotational speed estimated with the proper motions of the stars.
The large velocity dispersion with respect to the potential rotational motions suggests that the latter is not dominant in this system.

We also looked for additional indications of converging motions in the Taurus young cluster dynamics identified with {\it Gaia} \citep{esplinANDluhman2019,luhman2022}.
The bottom panel of Fig.~\ref{fig:prs3Ddustand13COwithYoungClusters} shows in detail the negative-slope pattern identified with the HOG in $n_{\rm H}$ and $^{12}$CO emission along with the radial velocities and distances to the young clusters. 
The roughly 3\,km\,s$^{-1}$ velocity difference across approximately 80\,pc along the LOS identified in the dust and $^{12}$CO is not followed by the young stellar clusters.
This discrepancy may be related to the current gas dynamics not dominating the young cluster dynamics.
Thus, it may indicate the different time scales that coexist within the MC, but does not contradict the compressive motions in the dust and gas.

The Taurus region with the most evident signature of converging motions identified with the HOG method corresponds to the southern portion of the B211/B213 region, which has been identified as an archetypical example of a star-forming filament \citep{schmalzl2010,palmeirim2013}.
The {\it Herschel} dust emission observations reveal the presence of striations perpendicular to the filament, generally oriented along the magnetic field direction as traced by optical polarization vectors \juan{and dust thermal emission} \citep{chapman2011,soler2019b,eswaraiah2021}.
These observations have led some authors to suggest that the material may be accreting along the striations onto the main filament \juan{\citep[see, for example,][and references therein]{pineda2022}}.
The typical velocities expected for the infalling material in this picture are around 1\,km\,s$^{-1}$, which is roughly compatible with our estimates, assuming that the dynamics that \cite{palmeirim2013} study on the plane of the sky are also those that we identified along the LOS.

In a follow-up to \cite{palmeirim2013}, \cite{shimajiri2019} studied the velocity pattern in the CO emission around the B211/B213 filament and identified the gas motions as the signature of ``mass accretion'' into that structure.
The authors conclusions are based in the emission at $v_{\rm LSR}$\,$\approx$\,6 to 7\,km\,s$^{-1}$ toward the northern-east filament edge and at $v_{\rm LSR}$\,$\approx$\,4 to 5\,km\,s$^{-1}$ toward the southern-west edge.
These velocities match the results of our analysis, but our analysis suggests that they are produced by the convergence of gas components along the line of sight rather than by gravitational collapse into the filament.

The fact that we found different dynamics in the stars and the combination of dust and gas is most likely related to the origin of the global contraction or compression of Taurus suggested by the HOG results.
If the stars, dust, and gas dynamics were dominated by the monolithic collapse toward the bottom of a gravitational potential well, we would expect coherent motion in all the tracers toward a particular location.
However, this is not what is obtained by the 3D motions of the stars in \cite{galli2019} or the results of the HOG.
Instead, we found compressive motions in the gas and dust in Taurus but not in the stars, as illustrated in Fig.~\ref{fig:prs3Ddustand13COwithYoungClusters}.

Our analysis suggests the convergence of gas components across the whole area of the MC, not continuous convergence in parts of the cloud.
This implies that the gas is brought together, rather than collapsing toward the local centers of gravitational collapse.
Thus, it is more likely that the convergence is being driven from the exterior of the MC, as with the injection of momentum from supernovae, rather than from the matter in its interior, as would be the case in gravitational collapse.
The former scenario gathers the gas while only marginally contributing to the motions of the stars. 
The latter influences the stars and the gas. 
Both scenarios can coexist in an MC.
Our results indicate that external compression is dominant in the Taurus MC.


\subsection{Potential origin of the Taurus MC}

The 3D dust maps used in this paper also enable precise constraints on the morphology of feedback-driven shells and superbubbles.
These structures appear as cavities in the interstellar medium and have been identified in the Taurus vicinity and other locations of the solar neighborhood \citep{pelgrims2020, bialy2021, marchal2022,foley2022}.

\begin{figure}[ht!]
\includegraphics[width=0.49\textwidth]{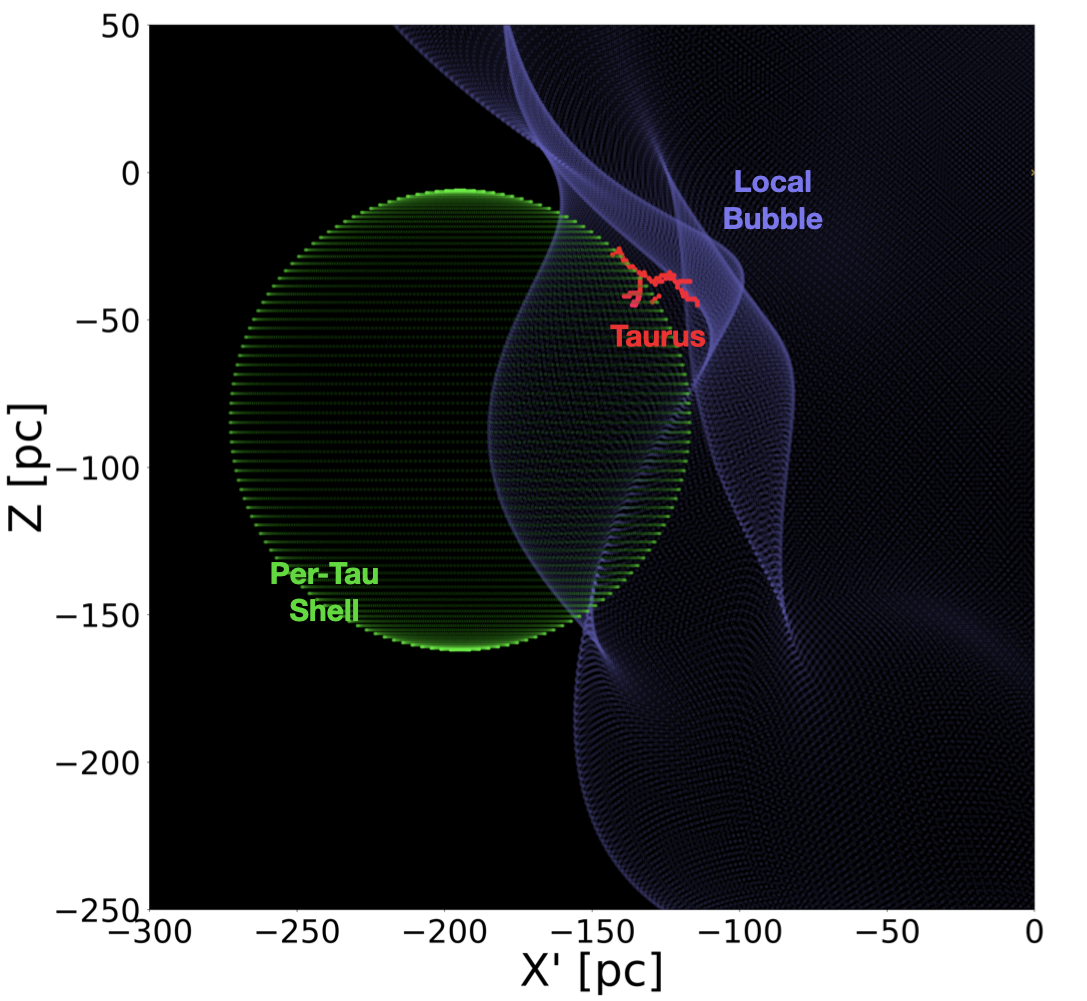}
\caption{Locations of the Taurus MC, Local Bubble, and the Per-Tau shell in an X-Z projection in heliocentric Galactic Cartesian coordinates.
The X-axis ($X'$) has been rotated anticlockwise by $\theta$\,$=$\,33\deg\ to show the best view of the intersection of the two shells.
The red points correspond to the 3D model of the dense gas in Taurus presented in \cite{zucker2021}, based on the 3D dust reconstruction presented in \cite{leike2020}. 
The Local Bubble surface, shown in purple, corresponds to the structure identified in \cite{pelgrims2020}, based on the 3D dust reconstruction presented in \cite{lallement2019}.
The Per-Tau shell surface, shown in green, corresponds to the structure identified in \cite{bialy2021}, based on the 3D dust reconstruction presented in \citet{leike2020}. 
Taurus lies at the intersection of the two surfaces.
}
\label{fig:bubbles}
\end{figure}

Figure~\ref{fig:bubbles} shows the relationship between the 3D position of the Taurus MC and the 3D morphology of the Per-Tau shell \citep{bialy2021} and the Local Bubble, whose boundary has been mapped out in \citet{pelgrims2020} using the 3D dust map of \citet{lallement2019}.
Taurus lies at the intersection of these two surfaces, where the Per-Tau shell seems to produce an indentation in the Local Bubble.
The near side of the cloud, which is touching the Local Bubble's surface, travels at higher radial velocities.
The far side of the cloud, which is touching the Per-Tau shell, travels at lower radial velocities.
Both of these observations strongly suggest that the relation between the radial velocity and distance revealed by the HOG in Fig.~\ref{fig:prs3Ddustand13COwithYoungClusters} is consistent with the formation of the Taurus MC as a result of large-scale gas compression via a ``snowplow" effect produced by the SN-driven expansion of two superbubbles \citep[see, for example,][]{inutsuka2015, dawson2013}.

The colliding shells scenario is consistent with the sheet-like morphology of Taurus in 3D dust at lower densities, as identified in \cite{zucker2021}.
The shock-compressed layers are possibly the origin of the network of filamentary structures embedded within the cloud at higher densities \citep{arzoumanian2013}.
These structures appear to evolve independently, as suggested by the proper motions of the stellar associations in Taurus \citep{roccatagliata2020}, and 
give rise to the young (a few megayears old) stellar population \citep[][]{luhman2018}.
Although the young stellar population does not rule out an MC origin by gravitational contraction, it matches what is expected if the MC had only formed recently due to the bubble collision, whereas the presence of an older population, for example, clusters with ages around 10\,Myr, would have been inconsistent with this scenario.

\section{Conclusions}\label{sec:conclusions}

We used the machine vision method HOG \citep{soler2019a} to study the relation between the 3D dust reconstruction presented in \cite{leike2020} and the H{\sc i}, $^{12}$CO, and $^{13}$CO emission toward the Taurus MC.
We showed that the morphological correlation quantified by the HOG provides a robust estimate of the distance to the MC by determining the dust parcel along the LOS that contributes the most to the observed distribution in integrated dust extinction or emission maps.
Thus, the HOG provides a powerful tool to determine distances to the features identified in 2D maps and disentangle the contributions of different dust parcels to the integrated maps.

We identified the anti-correlation between the mass traced by the 3D dust reconstruction and the H{\sc i} emission toward the Taurus MC.
We interpreted this result as the effect of the cold H{\sc i} associated with the MC.
We devote a follow-up paper to a detailed study of the H{\sc i} within Taurus.

We also found the potential signature of the converging motions in the HOG correlation between the 3D dust and $^{12}$CO.
The imprint of these motions is strongest toward the region around the star-forming filament B211/3.

The converging motion signature in the 3D dust and $^{12}$CO differs from the potential rotation signatures found in the proper motions of the stars in Taurus \citep{galli2019}.
However, this discrepancy is not in tension with our results for two main reasons.
First, the velocity of the potential rotational motion reported in \cite{galli2019} is below the 3D velocity dispersion of the stars, thus indicating that the rotation is not dominant in this system.
Second, the dynamics of the dust and gas are not necessarily the same as those of the stars.
The fact that we are finding converging motions in the gas and not in the stars can be interpreted as an indication that these are not dominated by a monolithic gravitational collapse, which would affect both the stars and the ISM, but rather by a large-scale shock that pushes and gathers the lower-density ISM.

The results of the HOG are consistent with the potential origin of the Taurus MC as the result of material compression by the expansion of the Local Bubble and the Per-Tau shell.
This dynamic imprint is a concrete observable that can be targeted in the analysis of numerical simulations and their associated synthetic observations to study the prevalence of this MC formation scenario. 
Also, as more superbubbles are mapped out in 3D across the solar neighborhood, the HOG technique presented here provides an opportunity to test MC formation mechanisms over a larger statistical sample.

We used the HOG method to map the distances revealed by dust into the radial velocity sampled by the gas tracers, revealing various aspects of one of the most studied MCs.
However, the correlation between gas tracers and 3D dust is just one of the MC characteristics that can be studied using the HOG and other novel statistical tools.
We devote two forthcoming papers to studying the \juan{cold atomic hydrogen} in and around the Taurus and its relation with the interstellar magnetic field.

\begin{acknowledgements}
The European Research Council funds JDS via the ERC Synergy Grant ``ECOGAL -- Understanding our Galactic ecosystem: From the disk of the Milky Way to the formation sites of stars and planets'' (project ID 855130).
CZ acknowledges support for this work through the National Aeronautics and Space Administration (NASA) Hubble Fellowship grant \#HST-HF2-51498.001 awarded by the Space Telescope Science Institute (STScI), operated by the Association of Universities for Research in Astronomy, Inc., for NASA, under contract NAS5-26555.
PFG acknowledges this research was conducted partly at the Jet Propulsion Laboratory, which the California Institute of Technology operates under NASA contract 80NM0018D0004.
RSK and SCOG acknowledge funding from the Deutsche Forschungsgemeinschaft (DFG) via the Collaborative Research Center (SFB 881, Project-ID 138713538) ``The Milky Way System'' (subprojects A1, B1, B2, and B8) and from the Heidelberg cluster of excellence (EXC 2181 - 390900948) ``STRUCTURES: A unifying approach to emergent phenomena in the physical world, mathematics, and complex data'', funded by the German Excellence Strategy.

\juan{We thank the anonymous referee for the thorough review and appreciate the suggestions that helped us improve this manuscript.}
Part of the crucial discussions that led to this work took part under the program Milky-Way-Gaia of the PSI2 project funded by the IDEX Paris-Saclay, ANR-11-IDEX-0003-02.
JDS thanks the following people who helped with their encouragement and conversation: Henrik Beuther, Jonas Syed, Robert Benjamin, Daniel Seifried, Rosine Lallement, Sara Rezaei Kh., and Sharon Meidt.
\juan{This research has made use of the SIMBAD database, operated at CDS, Strasbourg, France.}
The computations for this work were made at the Max-Planck Institute for Astronomy (MPIA) {\tt astro-node} servers.
{\it Software}: {\tt astropy} \citep{astropy2018}, {\tt SciPy} \citep{2020SciPy-NMeth}, {\tt astroHOG} \citep{astrohog2020}.
\end{acknowledgements}

\bibliographystyle{aa}
\bibliography{output.bbl}

\appendix

\section{Statistical significance of the HOG results}\label{appendix:significance}

The results of the HOG analysis are presented in terms of the projected Rayleigh statistic (V), presented in Eq.~\eqref{eq:prs} \citep[][]{jow2018}.
In the following sections, we evaluate the effect of the uncertainties in the 3D dust and emission maps and the impact of chance correlations.

\subsection{Error propagation}

The H{\sc i} and CO emission maps have uncertainties associated with the sensitivity of the instrument used to obtain them.
The 3D dust reconstruction has uncertainties related to reconstructing a continuous distribution based on discrete sampling.
We evaluated the effect of both uncertainties in the values of $V$ using Monte Carlo sampling.

\begin{figure*}[ht!]
\centerline{
\includegraphics[width=0.33\textwidth,angle=0,origin=c]{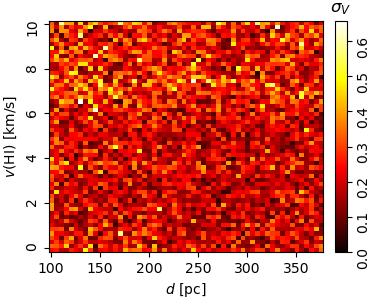}
\includegraphics[width=0.32\textwidth,angle=0,origin=c]{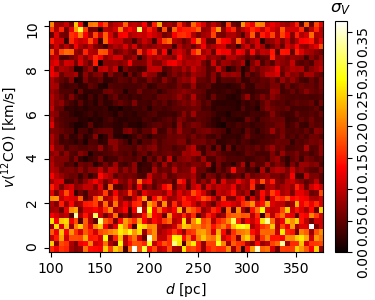}
\includegraphics[width=0.33\textwidth,angle=0,origin=c]{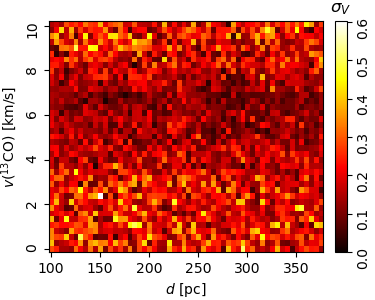}
}
\caption{Dispersion in the projected Rayleigh statistic ($V$) between the $n_{\rm H}$ and the H{\sc i}, $^{12}$CO, $^{13}$CO emission across distances and velocity channels estimated using Monte Carlo simulations of the input data.}
\label{fig:sigmaVplane}
\end{figure*}

For each velocity channel map and $n_{\rm H}$ slice, we generated ten Monte Carlo realizations with an amplitude equal to the 1-$\sigma$ uncertainty in each data set.
The results are 100 values of $V$ per distance and radial-velocity pair.
We computed the mean value $\left<V\right>$, which we reported in the main body of this paper, and the standard deviation around this value, $\sigma_{V}$, which we report in Fig.~\ref{fig:sigmaVplane}.

We found that the values of $\sigma_{V}$ are comparable to the 1-$\sigma$ absolute significance estimates for $V$ introduced in Sec.~\ref{sec:methods}.
The maximum values of $\sigma_{V}$ are around $V$\,$\approx$\,$1.64$, which correspond to the rejection of a uniform distribution of orientations between the gradients with a probability of 5\% \citep{batschelet1972}.
Thus, we conclude that the uncertainties in the emission PPV cubes and $n_{\rm H}$ are not systematically biasing the results of our analysis.  

\subsection{Impact of chance correlation}

Two independent 2D scalar fields, like the dust slices or gas emission velocity channels compared in this paper, can show some correlation due to the random alignment between their features, or chance correlation \citep[see, appendix~B in][]{soler2019a}.
We tested the impact of chance correlation in the values of $V$ using the same atomic and molecular emission PPV cubes but flipping the $n_{\rm H}$ cube.

\begin{figure*}[ht!]
\centerline{
\includegraphics[width=0.33\textwidth,angle=0,origin=c]{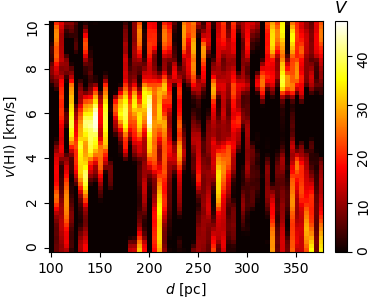}
\includegraphics[width=0.33\textwidth,angle=0,origin=c]{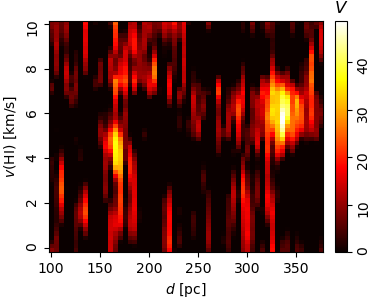}
\includegraphics[width=0.33\textwidth,angle=0,origin=c]{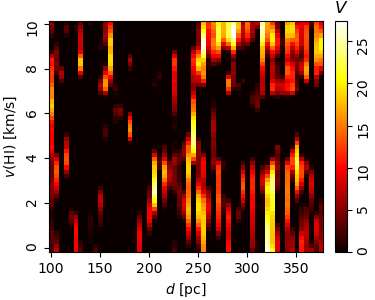}
}
\caption{\juan{Tests of change correlation between the} H{\sc i} emission and the 3D dust \juan{reconstruction for a $\Delta$\,$=$\,30\arcmin\ FWHM derivative kernel and a $I$\,$>$\,$\sigma_{I}$ mask}.
{\it Left}. Original maps. 
{\it Center}. Flipping \juan{the 3D dust cube} in the horizontal direction. 
{\it Right}. Flipping \juan{the 3D dust cube in the vertical and the horizontal directions}.   
}
\label{fig:jackknivesHIand3Ddust}
\end{figure*}

The results obtained for H{\sc i} and 3D dust, presented in Fig.~\ref{fig:jackknivesHIand3Ddust}, indicate a relatively high chance correlation between these two tracers.
The difference between the $V$ maximum values obtained with the original data and the two flipping tests are within 20\% of each other, thus setting tight constraints on the significance of the H{\sc i} to 3D dust correlation over the whole radial velocity and distance range.
However, the flipping tests confirm that the high $V$ values obtained around the distance of the Taurus MC are not the product of chance correlation.

\begin{figure*}[ht!]
\centerline{
\includegraphics[width=0.33\textwidth,angle=0,origin=c]{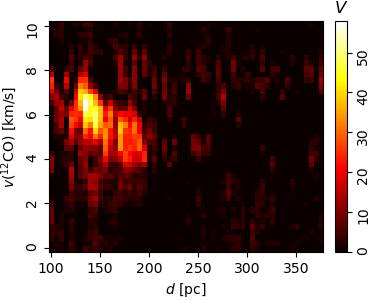}
\includegraphics[width=0.33\textwidth,angle=0,origin=c]{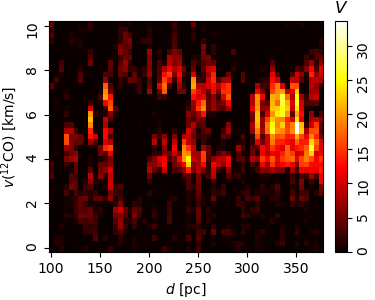}
\includegraphics[width=0.33\textwidth,angle=0,origin=c]{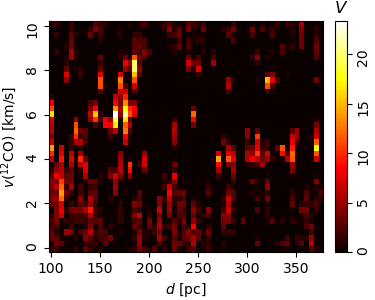}
}
\caption{Same as Fig.~\ref{fig:jackknivesHIand3Ddust}, but for $^{12}$CO.}
\label{fig:jackknives12COand3Ddust}
\end{figure*}

\begin{figure*}[ht!]
\centerline{
\includegraphics[width=0.33\textwidth,angle=0,origin=c]{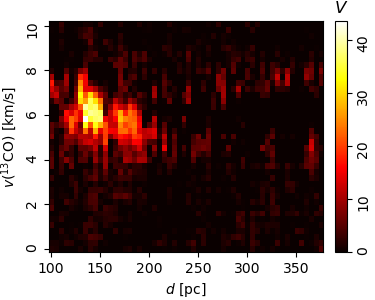}
\includegraphics[width=0.33\textwidth,angle=0,origin=c]{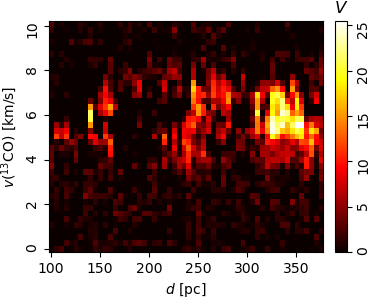}
\includegraphics[width=0.33\textwidth,angle=0,origin=c]{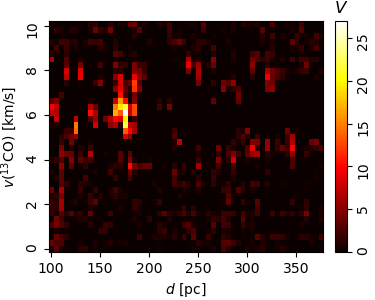}
}
\caption{Same as Fig.~\ref{fig:jackknivesHIand3Ddust}, but for $^{13}$CO.}
\label{fig:jackknives13COand3Ddust}
\end{figure*}

The results obtained for the two CO isotopologues and 3D dust, presented in Fig.~\ref{fig:jackknives12COand3Ddust} and Fig.~\ref{fig:jackknives13COand3Ddust}, show a lower chance correlation than in the H{\sc i} study.
This is expected given the more limited coverage of CO in the maps in relation to the almost ubiquitous H{\sc i}.
The maximum $V$ values in the original data are almost double those obtained in the flipping tests.
The $V$ distribution across distances and radial velocities is also very different in the flipping test. 
It demonstrates that the patterns interpreted in the main body of this paper are not the product of chance correlation.

\section{Angular resolution tests}\label{appendix:smoothing}

\juan{Throughout this paper, we have presented the results of the HOG analysis using the input observations at their native resolution.
This choice is justified because the computation of the gradients employing the Gaussian derivatives convolves the images to the common scale set by the size of the derivative kernel.
There is, however, an increase in the signal-to-noise ratio ($S/N$) if the input data is convolved to a common angular resolution larger than the beam size before computing the HOG.
This change in $S/N$ potentially produces different results of the HOG analysis, either by modifying the low-signal regions that are masked or by setting a different mean in the Monte Carlo realizations.
In this appendix, we present the results obtained when smoothing the maps to a common angular resolution before computing the HOG. 
}

\begin{figure*}[ht!]
\centerline{
\includegraphics[width=0.33\textwidth,angle=0,origin=c]{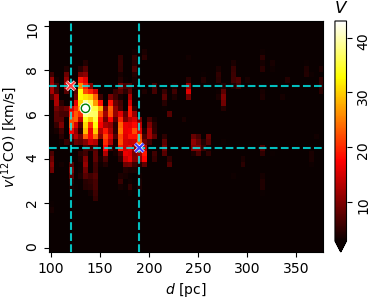}
\includegraphics[width=0.33\textwidth,angle=0,origin=c]{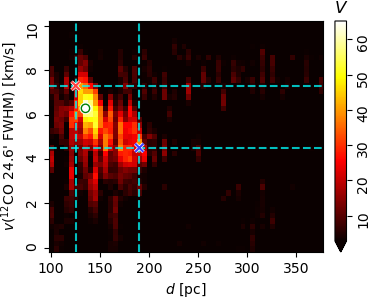}
\includegraphics[width=0.33\textwidth,angle=0,origin=c]{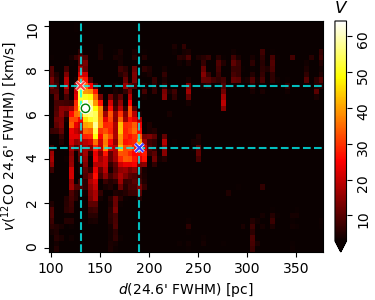}
}
\caption{\juan{Correlation between the distribution of hydrogen volume density ($n_{\rm H}$) derived from the \cite{leike2020} 3D dust reconstruction and $^{12}$CO emission across distances and radial velocities as quantified by the projected Rayleigh statistic ($V$) using a 30\parcm0 FWHM derivative kernel.
The three panels correspond to three different input data as follows.
{\it Left}. 3D dust and $^{12}$CO at their native angular resolution, 6\parcm87 and 45\arcsec, respectively. 
{\it Center}. 3D dust at its native angular resolution and $^{12}$CO smoothed to a 24\parcm6 FWHM resolution.
{\it Right}. 3D dust and $^{12}$CO smoothed to a 24\parcm6 FWHM resolution.
In all three cases, regions with $S/N$\,$<$\,$3$ are excluded from the analysis.}
}
\label{fig:SmoothingTestMask}
\end{figure*}

\juan{Figure~\ref{fig:SmoothingTestMask} presents the results of the HOG analysis for the $^{12}$CO emission and the 3D dust using three different combinations of the input maps: one in which both have their native angular resolution, one in which the 12CO is smoothed with a 24\parcm6 FWHM kernel, and one in which both maps are smoothed with this kernel. 
The 24\parcm6 kernel width corresponds to 1\,pc at the mean distance to Taurus.
In the HOG computation, we employed a 30\parcm0 FWHM derivative kernel and a $S/N$\,$>$\,$3$ mask.
The smoothing of the $^{12}$CO input data produces higher maximum values of $V$ and high $V$ in distance and velocity channel pairs that are not statistically significant with the raw input data.
This is most likely due to the reduction of the masked regions in the smoothed maps due to the increase in $S/N$ for the lower-resolution data.
The 3$\varsigma_{V}$ distance and velocity ranges are identical in the three cases.
The changes in the distribution of $V$ across distance and velocity channels do not significantly affect the main trend reported in the paper.}
 
\begin{figure*}[ht!]
\centerline{
\includegraphics[width=0.33\textwidth,angle=0,origin=c]{{taurusHOG_12COand3DdustLITEd5.0_ksz1800.0_nruns4_v0.0to10.0_d100.0to375.0_sigmalim0.0_nruns4_VplaneWithLines}.png}
\includegraphics[width=0.33\textwidth,angle=0,origin=c]{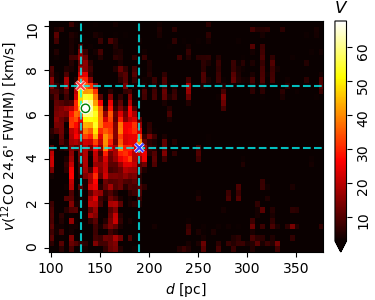}
\includegraphics[width=0.33\textwidth,angle=0,origin=c]{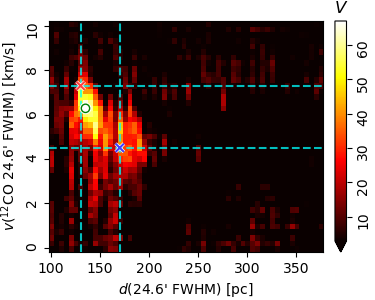}
}
\caption{\juan{Same as Fig.~\ref{fig:CollapseTestVplane}, but for Monte Carlo realizations of input maps instead of masking.}}
\label{fig:SmoothingTestMC}
\end{figure*}

\juan{Figure~\ref{fig:SmoothingTestMC} presents the results of the HOG analysis for the $^{12}$CO emission and the 3D dust using the same three different combinations of the input map smoothing, but with the HOG computed using the Monte Carlo realizations introduced in App.~\ref{appendix:significance} instead of masking.
The results of this test indicate an increase in $V$ of around 10\%, but the same global distribution of $V$ across distance and velocity channels in the three cases.
This confirms that the differences found between the three combinations of input maps in Fig.~\ref{fig:SmoothingTestMask} are produced by the masking.
The 3$\sigma_{V}$ distance and velocity ranges are essentially identical in the three cases, just with a slight change in the furthest distance channel due to the general increase in the $V$.
However, the global trend within the 3$\sigma_{V}$ range is unchanged. 
These results indicate that the initial smoothing does not significantly affect the results of our analysis and supports our choice of Monte Carlo realizations instead of masking.
}

\section{Masking tests}\label{appendix:masking}

\begin{figure*}[ht!]
\centerline{
\includegraphics[width=0.33\textwidth,angle=0,origin=c]{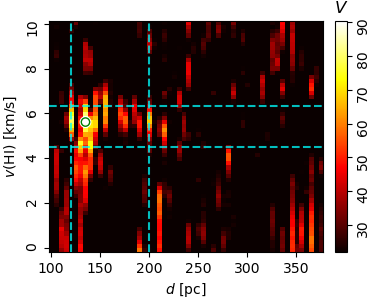}
\includegraphics[width=0.33\textwidth,angle=0,origin=c]{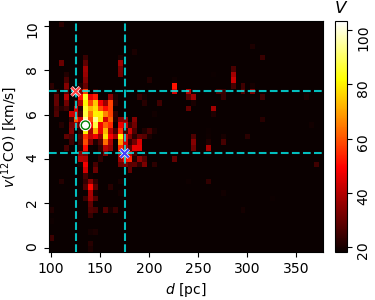}
\includegraphics[width=0.33\textwidth,angle=0,origin=c]{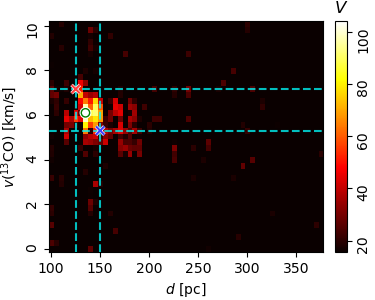}
}
\caption{Same as Fig.~\ref{fig:CollapseTestVplane}, but masking regions where $A_{\rm V}$\,$>$\,1.0.}
\label{fig:CollapseTestVplaneAVmasks1}
\end{figure*}

\begin{figure*}[ht!]
\centerline{
\includegraphics[width=0.33\textwidth,angle=0,origin=c]{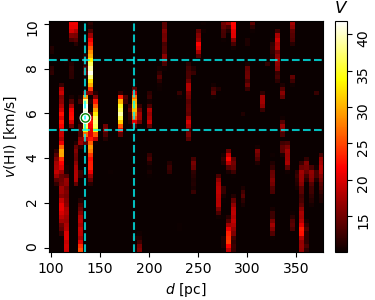}
\includegraphics[width=0.33\textwidth,angle=0,origin=c]{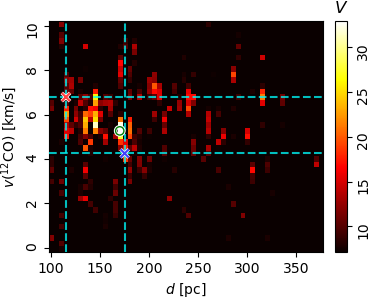}
\includegraphics[width=0.33\textwidth,angle=0,origin=c]{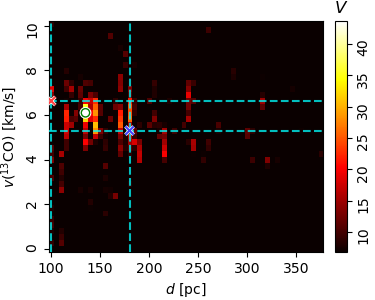}
}
\caption{Same as Fig.~\ref{fig:CollapseTestVplane}, but masking regions where $A_{\rm V}$\,$>$\,3.0.}
\label{fig:CollapseTestVplaneAVmasks3}
\end{figure*}

Traditionally, the Taurus MC was defined through the visual extinction $A_{\rm V}$ \citep[see, for example,][]{cambresy1999}.
Here, we discuss the results obtained when applying the HOG only to the portions of the Taurus MC within the regions defined by cuts in $A_{\rm V}$.
Fig.~\ref{fig:CollapseTestVplaneAVmasks1} shows that the optically thick regions in the $V$ band, $A_{\rm V}$\,$>$\,1.0, maintain the trends described in the main body of this paper, which are the high correlation between H{\sc i} and 3D dust at the distance of Taurus and the high correlation between $^{12}$CO and 3D implying that the near side of the cloud is moving at higher velocities and the far side is moving at lower velocities.
The same patterns are lost with a higher extinction cut, $A_{\rm V}$\,$>$\,3.0, as shown in Fig.~\ref{fig:CollapseTestVplaneAVmasks3}.
The main reason for this loss in signal is the lack of independent gradient vectors, which is reflected in the lower $V$ values.
At this relatively high extinction, the 3D dust reconstruction also saturates its density values, and the gradients used to trace its correlation with the gas emission are less meaningful.

\section{Block averaging}\label{appendix:blocks}

\begin{figure*}[ht!]
\centerline{
\includegraphics[width=0.99\textwidth,angle=0,origin=c]{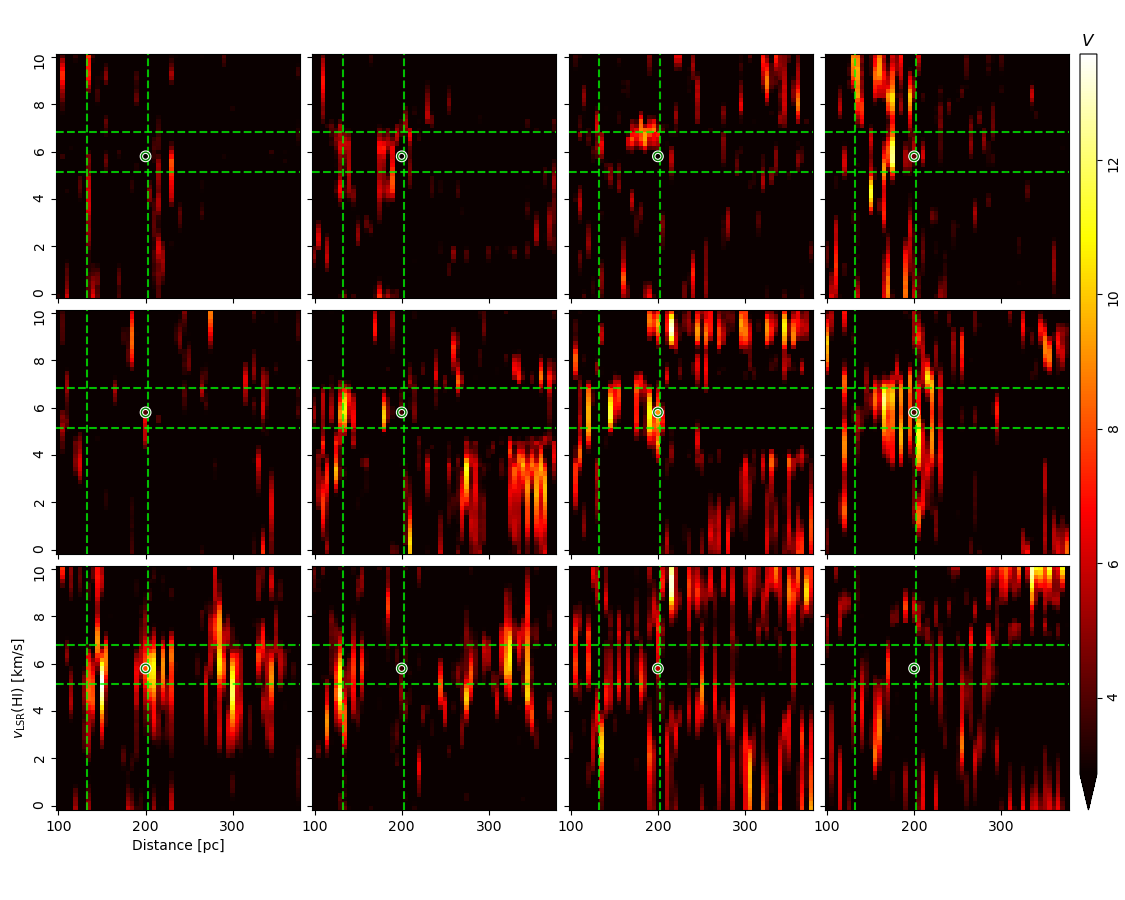}
}
\caption{Same as Fig.~\ref{fig:blockaverageVplane12CO}, but for H{\sc i}.}
\label{fig:blockaverageVplaneHI}
\end{figure*}

\begin{figure*}[ht!]
\centerline{
\includegraphics[width=0.99\textwidth,angle=0,origin=c]{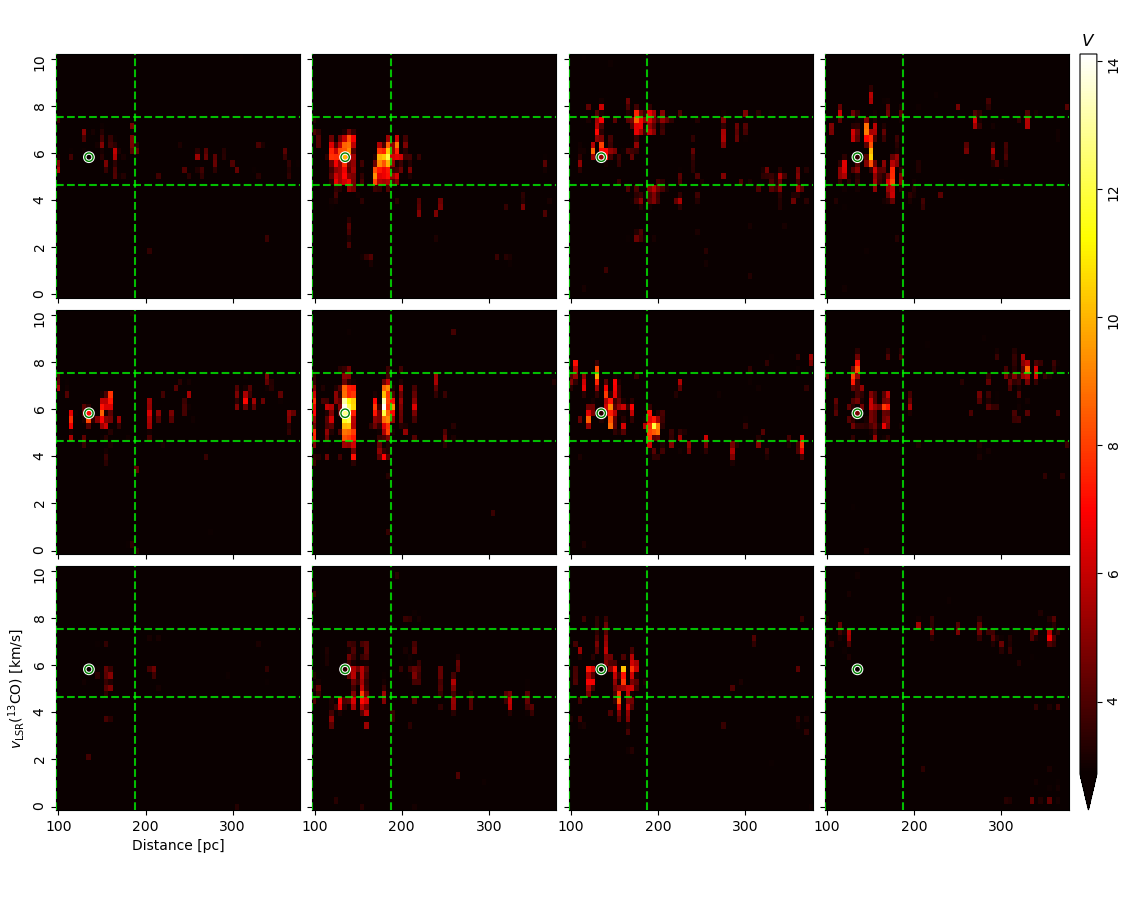}
}
\caption{Same as Fig.~\ref{fig:blockaverageVplane12CO}, but for $^{13}$CO.}
\label{fig:blockaverageVplane13CO}
\end{figure*}

For the sake of completeness, we report the results of the HOG comparison between 3D dust and H{\sc i} and $^{13}$CO in 4\,$\times$\,3 blocks across the Taurus MCs in Fig.~\ref{fig:blockaverageVplaneHI} and Fig.~\ref{fig:blockaverageVplane13CO}, respectively.
In the case of $^{13}$CO and 3D dust, the panel corresponding to the vicinity of the B211/B213 region
displays the pattern linking the near side of the MC moving at higher velocities and the far side moving at lower velocities that was reported for $^{12}$CO and 3D dust in the main body of the paper.
However, the levels of significance are low and limit any further interpretation.

\section{Test using 3D extinction maps }\label{appendix:extinctionmaps}

To contrast the results obtained with the $n_{\rm H}$ reconstructions, we also applied the HOG analysis to the 3D maps of dust reddening, $E(B-V)$, based on Gaia parallaxes and stellar photometry from Pan-STARRS 1 and 2MASS \citep{green2019}.
The main difference between the 3D extinction maps and the 3D dust reconstructions is that the former corresponds to a cumulative quantity, which saturates to a maximum value.
They also do not account for the spatial correlations between contiguous lines of sight. 
Still, we highlight the importance of the 3D density reconstruction for our results by computing the HOG with the 3D extinction maps toward Taurus. 

\begin{figure*}[ht!]
\centerline{
\includegraphics[width=0.33\textwidth,angle=0,origin=c]{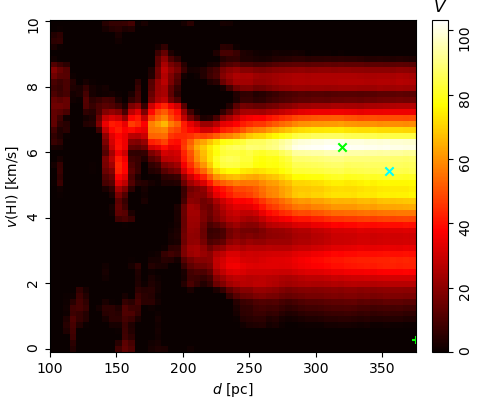}
\includegraphics[width=0.33\textwidth,angle=0,origin=c]{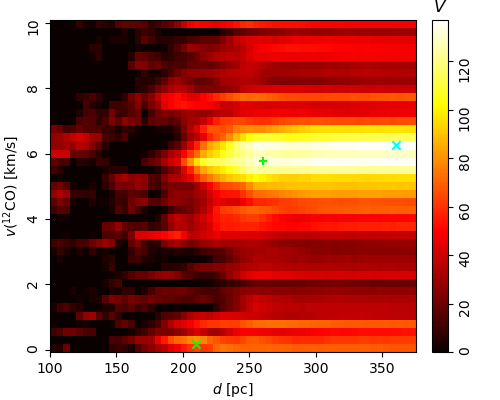}
\includegraphics[width=0.33\textwidth,angle=0,origin=c]{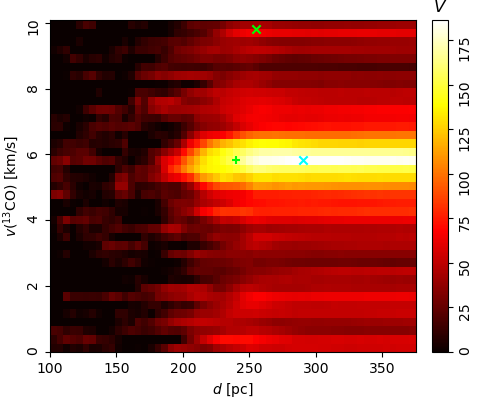}
}
\caption{Correlation between the dust extinction across distances \citep{green2019} and H{\sc i} emission across distances and velocity channels as quantified by the projected Rayleigh statistic ($V$) and the Pearson correlation coefficient ($r_{\rm Pearson}$).}
\label{fig:VplaneHIandLOSextintion}
\end{figure*}

Figure~\ref{fig:VplaneHIandLOSextintion} shows the results of the correlation between the H{\sc i}, $^{12}$CO, and $^{13}$CO emission and the dust reddening along the line of sight toward the Taurus MC.
The HOG method consistently recovers the spatial correlation between the gas emission and the extinction, but the distance where the increase in $V$ indicates the presence of the MC is larger than that of Taurus.
This fact is related to the lack of spatial correlation for contiguous pixels in the extinction maps and the cumulative nature of extinction: a meaningful gradient only appears when enough contiguous pixels are correlated by the increase in the number of background stars. 
The cumulative nature of the 3D extinction maps also implies that they are not sensitive to the structure behind the dominant dust feature in the map, which can only be recovered as a differential quantity between extinction slabs around the line of sight.
Thus, the results reported in the main body of this paper are only possible thanks to the reconstruction of a smooth distribution both in the plane of the sky and along the line of sight in the 3D density maps.

\end{document}